\documentclass[twocolumn,english,aps,superscriptaddress,prb]{revtex4-1}

\usepackage{babel}
\usepackage{amsmath}
\usepackage{amssymb}
\usepackage{graphicx}
\usepackage{stmaryrd}
\usepackage{physics}
\usepackage{bm}
\usepackage[colorlinks, linkcolor=mycol, citecolor=mycol, urlcolor=mycol, breaklinks]{hyperref}
\usepackage{braket}
\usepackage{bbold}
\usepackage{bm}
\usepackage{ulem}
\usepackage{tabularx, colortbl}
\usepackage{comment}
\usepackage{appendix}
\usepackage{young}
\usepackage{ytableau}
\usepackage{epstopdf}
\usepackage{xcolor}
\usepackage{tikz}
\usetikzlibrary{decorations}
\usetikzlibrary{decorations.pathmorphing}
\usetikzlibrary{decorations.pathreplacing}
\usetikzlibrary{decorations.markings}
\usetikzlibrary{shapes.misc}
\usetikzlibrary{calc}

\hypersetup{
    colorlinks=true,
    linkcolor=blue,
    citecolor=blue,    
    urlcolor=blue,
}

\definecolor{mycol}{RGB}{10,55,130}

\begin{document}

\title{Exact Diagonalization of $\mathrm{SU}(N)$ Fermi-Hubbard Models}

\author{Thomas Botzung}
\affiliation{Laboratoire de Physique et Mod\'elisation des Milieux Condens\'es, Universit\'e Grenoble Alpes and CNRS, 25 avenue des Martyrs, 38042 Grenoble, France}

\author{Pierre Nataf}
\affiliation{Laboratoire de Physique et Mod\'elisation des Milieux Condens\'es, Universit\'e Grenoble Alpes and CNRS, 25 avenue des Martyrs, 38042 Grenoble, France}

\date{\today}
\begin{abstract} 
We show how to perform exact diagonalizations of $\mathrm{SU}(N)$ Fermi-Hubbard models on $L$-site clusters separately in each irreducible representation ({irrep}) of $\mathrm{SU}(N)$. 
Using the representation theory of the unitary group $\mathrm{U}(L)$, we demonstrate that a convenient orthonormal basis, on which matrix elements of the Hamiltonian are very simple,  
 is given by the set of {\it semistandard Young tableaux}  (or, equivalently the Gelfand-Tsetlin patterns) corresponding to the targeted irrep. As an application of this color factorization, we study the robustness of some  $\mathrm{SU}(N)$ phases predicted in the Heisenberg limit upon decreasing the on-site interaction $U$ on various lattices of size $L \leq 12$ and for $2 \leq N \leq 6$. In particular, we show that a long-range color ordered phase emerges for intermediate $U$
for $N=4$ at filling $1/4$ on the triangular lattice.

\end{abstract}

\maketitle



The Fermi-Hubbard model (FHM) is among the most important models in condensed matter~\cite{Hubbard_1963,Gutzwiller_1963, Scalapino_2012}. In particular, the $\mathrm{SU}(2)$ FHM on the square lattice might describe the physics of electrons carrying a spin one half in cuprate superconductors~\cite{Anderson_1987, Rice_1988,Scalapino_2012} and has motivated numerous theoretical investigations~\cite{review_Arovas_2022,review_Corboz_2022}. 
Considering $N$, the number of degenerate orbitals, as an integer parameter of the models,  a natural extension of the $\mathrm{SU}(N=2)$ FHM is the $\mathrm{SU}(N)$ FHM~\cite{Assaraf_1999,honerkamp_ultrcold_2004,capponi_phases_2016}. 

This higher symmetry group Hamiltonian, first introduced as a theoretical tool to provide an asymptotic description of spins $1/2$ in the large $N$ limit~\cite{affleck_exact_1986,Affleck_1988,Rokhsar_1990,Marder_1990}, can also describe some condensed matter systems like transition metal compounds~\cite{Khomskii_1998,Yamada_2018} or graphene with $\mathrm{SU}(4)$  spin valley symmetry or in twisted bilayer~\cite{zhang2021}.
Alternatively, alkaline-earth cold atoms like $^{173}$Yb or $^{87}$Sr can simulate $\mathrm{SU}(N)$-invariant FHMs for $N$ up to $10$ on various engineered optical lattices~\cite{wu_exact_2003,Wu_review_2006,gorshkov_two_2010,Cazalilla_2014}.
Additionally, the continuous experimental achievements in this field~\cite{taie_su6_2012,hofrichter_direct_2016,Becker_2021,taie2020observation,Fallani_2022,pasqualetti2023equation} brought theoreticians to investigate these systems in order to look for exotic phases that would generalize their $N=2$ counterpart. %

Apart from some quantum Monte Carlo (QMC) studies \cite{Wang_2014,Xu_2018} addressing the $\mathrm{SU}(N)$ FHM at half filling for a wide range of positive on-site interaction $U$, most of the theoretical investigations focused on the large $U$ limit, where the atoms, in the Mott insulating phase, are described by $\mathrm{SU}(N)$ Heisenberg models (HMs)~\cite{Read_1989,Zhang_1998,Mila_2003,greiter_2007,Corboz_2011,Bauer_2012,Quella_2014,Weichselbaum_2018}.
Depending on the lattices and on the number of colors $N$, different two-dimensional phases are predicted  at $T=0$, among which the $\mathrm{SU}(N)$ plaquette phases~\cite{Paramekanti_2007,Corboz_2012,Assad_2013,Nataf_honey_2016,Gauthe_2021} are cousins of the valence bond states for spins $1/2$, the N\'eel long-range color ordered (LRO) states~\cite{Corboz_2011,Bauer_2012} are analogous to the famous ($\pi,\pi$) (respectively, $120^{\circ}$) N\'eel states existing on the square~\cite{Anderson_1952,Manousakis_1991} (respectively. the triangular~\cite{Lecheminant_1994,Sorella_1999}) lattice for $N=2$, and diverse kinds of $\mathrm{SU}(N)$ spin liquids~\cite{Wang_2009,Hermele_2011,Corboz_prx_2012,Lai_2013,Chen_Hazzard_2016,Nataf_chiral_2016,Boos_2020,Keselman_2020,Chen_2021, Jin_2022} generalizing the Anderson resonating valence bond states~\cite{Anderson1973ResonatingVB,Anderson_1987}.
 Except for the one-dimensional system where there is a Bethe ansatz solution~\cite{sutherland}, the theoretical investigation of these models, based on advanced numerical tools, is challenging mainly because the dimension of the full Hilbert space on finite-size lattices increases
 exponentially, being equal to $N^L$, where $L$ is the number of sites of the cluster,  for filling $1/N$ (exactly one particle per site) in the $\mathrm{SU}(N)$  HM.
 
 However, it was realized that working in the $\mathrm{SU}(N)$ singlet subspace, which usually contains the ground state (GS) in the antiferromagnetic case, is very advantageous as its dimension 
 is much smaller than $N^L$. For instance for $N=6$ and $L=12$, such a dimension is equal to $132$, while $N^L \equiv 6^{12}\approx 2 \times 10^9$.
 \,In addition, the exact diagonalization (ED) of the HM directly in the $\mathrm{SU}(N)$ singlet subspace can be made easy on the basis of standard Young Tableaux  (SYT) using the orthogonal representation of the group of permutations $\mathcal{S}_L$~\cite{nataf_exact_2014}. 
 It is crucial to extend this theory to the $\mathrm{SU}(N)$ FHM as the dimension of the full Hilbert space is even larger, i.e equal to $2^{NL}$.
In fact, for $N=6$, $L=12$ at filling $1/6$ the dimension of the singlet subspace for the $\mathrm{SU}(6)$ FHM is  $\approx 14 \times 10^6$ while the full Hilbert space has dimension $2^{72} \approx 5 \times 10^{21}$.

In this Letter, we use the representation theory of the Lie group $\mathrm{U}(L)$ to show how to perform ED of the $\mathrm{SU}(N)$ FHM directly in each irreducible representation (irrep). After the description of the method,
we apply the procedure to show some ED results on square and triangular clusters of size up to $L=12$ and for $N$ up to $N=6$ to see how robust are some $\mathrm{SU}(N)$ Mott insulating phases while decreasing the on-site repulsion.

The Hamiltonian for the $\mathrm{SU}(N)$ FHM reads
\begin{equation}
\label{Hamiltonian}
H= \sum_{\langle i,j \rangle}  \Big{(}  -t_{ij} E_{ij}+ \text{H.c.} \Big{)} + \frac{U}{2}  \sum_{i=1}^L E_{ii}^2, 
\end{equation}
where the $t_{ij}$ are the (possibly complex) hopping amplitude between sites $i$ and $j$ of a $L$-site finite cluster, and the on-site interaction amplitude is $U$. 
The $\mathrm{SU}(N)$-invariant hopping terms $E_{ij}=E_{ji}^{\dag}=\sum_{\sigma=1}^N c^{\dag}_{i \sigma}c_{j \sigma}$ satisfy the commutation relation of the $\mathrm{U}(L)$ generators ($\forall \, {1 \leq i,j,k,l \leq L}$),
\begin{equation}
\label{commutation}
[E_{ij},E_{kl}]=\delta_{jk}E_{il}-\delta_{li}E_{kj}, 
\end{equation}
so that the Hamiltonian in Eq. \eqref{Hamiltonian}, where the integer parameter $N$ is {\it hidden}, can be seen as an element of the Lie algebra of the unitary group $\mathrm{U}(L)$ \footnote{Or, more precisely, of its universal enveloping algebra \cite{dixmier1977enveloping}, since it contains some square of generators}.
It should be considered as the counterpart of the quantum permutation Hamiltonian for the $\mathrm{SU}(N)$-invariant HM, 
i.e., $H=\sum_{\langle i,j \rangle} J_{ij} P_{ij} + \text{H.c.} $, with  $P_{ij}$ (respectively, $J_{ij}$  ) the permutation (respectively, coupling constant) between interacting sites $i$ and $j$,
 for which the representation theory of the algebra of the group of permutations was used to perform ED directly and separately in each irrep of $\mathrm{SU}(N)$~\cite{nataf_exact_2014,wan_exact_2017}.
 
\begin{figure}[h!]
\includegraphics[width=0.5\textwidth]{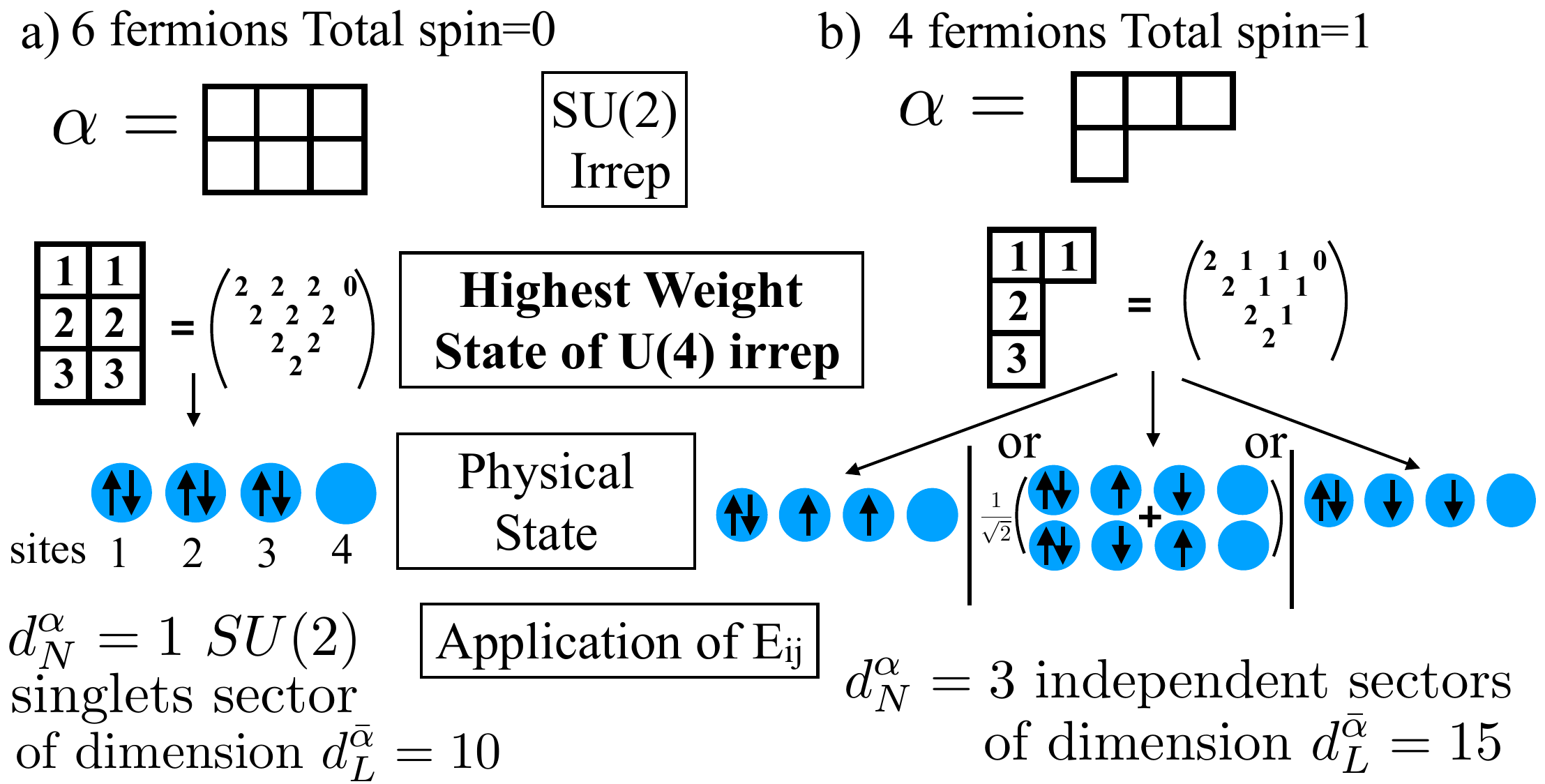}
\caption{Examples of $\mathrm{SU}(N=2)$ irrep $\alpha$ for a $L=4$-site cluster and $M=6$ fermions in (a) and $M=4$ fermions in b).
The Young tableaux are $\alpha=[3,3]$ ( respectively, $\alpha=[3,1]$ ), representing singlets (respectively, spin 1) states.
We associate the transpose YT, flipping the shape: $\alpha=[3,3] \rightarrow \bar{\alpha}=[2,2,2]$ for (a) and $\alpha=[3,1] \rightarrow \bar{\alpha}=[2,1,1]$ for (b).
We fill up $\bar{\alpha}$ to get the highest weight state ($\vert \text{hws} \rangle$) to which we associate $d^{\alpha}_{N=2}$ physical states, each of which generates
an independent $d^{\bar{\alpha}}_L$-dimensional sector invariant under the application of the operators $E_{ij}$. See text for details.}
\label{fig1}
\end{figure}

 We remind the reader that an irrep of $\mathrm{SU}(N)$ 
  is labeled by a Young Tableau (YT), or shape $\boldmath{\alpha}$ (see Fig. \ref{fig1}), the $N$ rows of which represent $N$ integers $\alpha=[\alpha_1,\alpha_2,\dots,\alpha_N]$ such that $\alpha_1\geq \alpha_2 \geq\dots\geq \alpha_N\geq0$ and $\sum_{i=1}^N \alpha_i=M$, where $M$ is the number of particles [i.e.,  the filling is $M/(LN)$].
  Calling $\mathcal{H}^{M,N}_L$, the Hilbert space for $L$ sites and $M$  $\mathrm{SU}(N)$ fermions, its
   dimension $ D^{M,N}_L \equiv \text{dim} (\mathcal{H}^{M,N}_L)$ is \cite{supp_mat}
  \begin{equation}
\label{decomposition}
\sum_{\boldmath{\alpha}} h^{\bar{\alpha}}_L  \prod_{i=1}^L \binom{N}{\bar{\alpha}_i}=\sum_{\boldmath{\alpha}} d^{\alpha}_N  d^{\bar{\alpha}}_L,
\end{equation}
where the sums on both sides run over all the YT $\alpha$ of $M$ boxes, with maximum $L$ columns and $N$ rows.
$\bar{\alpha}=[\bar{\alpha}_1,\dots,\bar{\alpha}_L]$ is defined as the {\it transpose}  YT of $\alpha$, transforming rows into columns (cf.  Fig.~\ref{fig1} for some examples).
On the lhs of Eq.~\eqref{decomposition}, $\bar{\alpha}$ is a 
 distribution of fermions: $\bar{\alpha}_j$ being the number of fermions (necessarily $\leq N$) on site $j$ for $1 \leq j \leq L$;
$ \prod_{i=1}^L \binom{N}{\bar{\alpha}_i}$ is the number of states for such a distribution. 
The factor  $h^{\bar{\alpha}}_L$, defined as $h^{\bar{\alpha}}_L=L!/ \prod_{k=0}^N (n_k^{\bar{\alpha}})!$, where $n_k^{\bar{\alpha}}=\text{Cardinal}\{j \in \llbracket 1;L \rrbracket : \bar{\alpha}_j=k\}$, is the number of distributions corresponding to
a given partition $\bar{\alpha}$, while permuting the $\bar{\alpha}_j$ (or the site indices $j$) for $1 \leq j \leq L$. In the rhs of Eq.~\eqref{decomposition}, $d^{\alpha}_N $ (respectively, $d^{\bar{\alpha}}_L$) stands for the dimension of the $\mathrm{SU}(N)$ irrep $\alpha$ [respectively, the $\mathrm{U}(L)$ irrep $\bar{\alpha}$]\footnote{The irreps of $\mathrm{U}(N)$ or $\mathrm{SU}(N)$ are basically the same \cite{supp_mat}; distinction is relevant when we include/exclude representation of generators with non vanishing trace like some combinations of $E_{ii}$.}, which we can calculate using existing  formulas ~\cite{supp_mat}.
These dimensions are equal to the number of semistandard Young tableaux (ssYT) of shape $\alpha$ (respectively, $\bar{\alpha}$) filled with numbers from $1$ to $N$ (respectively, $L$), since these latter form a basis of the $\mathrm{SU}(N)$ or $\mathrm{U}(L)$ irrep.
Given a $\mathrm{U}(L)$ irrep represented by some YT, a ssYT is filled up with integer numbers from 1 to L in nondescending order from left to right in any row (repetitions allowed), and in strictly ascending order (repetitions not allowed) from top to bottom in any column [cf. Fig.~\ref{fig1} and Eq.~\eqref{eq_ssYT} for some examples].

As detailed below and in the Supplemental Material \cite{supp_mat}, for the color-invariant $\mathrm{SU}(N)$ FHM [cf.~Eq.~\eqref{Hamiltonian}], the Hilbert space $\mathcal{H}^{M,N}_L$ can be decomposed, or {\it color-factorized}, following the rhs of the equation for the dimension $D^{M,N}_L$, i.e.,  Eq.~\eqref{decomposition}. In particular, targeting a given collective $\mathrm{SU}(N)$ irrep $\alpha$, we will need to diagonalize a matrix of dimension $d^{\bar{\alpha}}_L$, and there will be $d^{\alpha}_N$ independent copies (some multiplicity) of the corresponding spectrum in the full energy spectrum of the model.  For instance, when $M$ is a multiple of $N$, one important sector is the $\mathrm{SU}(N)$ singlet sector, as it usually contains the lowest energy eigenstates (in the large $U>0$ limit, for instance); it is labeled by the perfectly rectangular $N-$rows YT $\alpha=\alpha_{\mathcal{S},M}\equiv [M/N,M/N,\dots,M/N]$. In this case, $d^{\alpha_{\mathcal{S},M}}_N=1$ and for $L=12$ at filling $1/N$ ($M=L$), one has for instance,
$d^{\bar{\alpha}_{\mathcal{S},M=L}}_{L=12}=13 \, 026 \, 013 $ for $N=4$ and $d^{\bar{\alpha}_{\mathcal{S},M=L}}_{L=12}=14 \, 158 \, 144$ for $N=6$.
This should be compared to the dimensions of the sector usually addressed in standard ED with a fixed number of fermions of each color [conserving the U($1$) symmetry], which is  $d^{U(1)}_{L,M,N}= \binom{L}{M/N}^N$: one has $d^{U(1)}_{L=M=12,N=4}\approx 2.34 \times10^9$ (respectively,  $d^{U(1)}_{L=M=12,N=6}\approx 8.27 \times 10^{10}$). As $N$ increases, it is more and more advantageous to implement the full $\mathrm{SU}(N)$ symmetry, working in the $\mathrm{SU}(N)$ singlet sector and, more generally, in a sector of a given irrep $\alpha$.

For a given $\mathrm{U}(L)$ irrep $\bar{\alpha}$, the {\it highest weight state} ($\vert \text{hws} \rangle $), uniquely (up to some similarity) and fully determines the irrep as one can generate the entire basis by applications of the generators  $E_{ij}$  (for $1 \leq i,j \leq L$).
The $\vert \text{hws} \rangle $ is represented by the shape $\bar{\alpha}$ filled with $1$ for the first row (of length $\bar{\alpha}_1$), $2$ for the second row of length $\bar{\alpha}_2$), etc.(cf. Fig. \ref{fig1}).
It is defined by the following properties: $E_{ii} \vert \text{hws} \rangle=\bar{\alpha}_i \vert \text{hws} \rangle $  $\forall \, i \in \llbracket 1;L \rrbracket$  and  $E_{ij} \vert \text{hws} \rangle=0$ for $i<j$ \cite{Paldus_2021}.
Crucially, in  $\mathcal{H}^{M,N}_L$, there are $d^{\alpha}_N$ orthonormal states $\vert \phi^{\text{hws}}_{\alpha, k}\rangle$ ($k=1, \dots, d^{\alpha}_N$) which have these properties and can then be represented by the same ssYT associated with the $\vert \text{hws} \rangle$. For example, for the $\mathrm{SU}(N)$ singlet irrep $\alpha_{\mathcal{S},M}$, there is only one state $\vert \phi^{\text{hws}}_{\alpha, 1}\rangle$ and it is the product of $\mathrm{SU}(N)$ singlets for sites $1, 2, \dots, M/N$, with no particles on sites $M/N+1, \dots, L$ [cf. Fig. \ref{fig1} in the Supplemental Material and \cite{supp_mat} for details about $\mathrm{SU}(N)$ singlets].

On the basis of the ssYT, which are equivalent to the Gelfand-Tsetlin (GT) patterns~\cite{alex_2011,supp_mat}, the matrix elements of the infinitesimal generators $E_{p p}$, which are the occupation numbers on site $p$ for $p=1, \dots, L$ and of $E_{p-1 p}$ (respectively, $E_{p p-1}$ ), which generalize the lowering operator $J_{-}$ (respectively, $J_{+}$) for $\mathrm{U}(2)$, are very simple.
Found by Gelfand and Tsetlin~\cite{Gelfand_1950}, we detail them in the Appendix.
As an illustrative example, for the $\mathrm{SU}(4)$ adjoint irrep at filling $1/4$ for $L=12$ (the basis has then $57 \, 972 \,915$ elements), we have, for instance,
 \begin{align}
\label{eq_ssYT}
 \ytableausetup{smalltableaux}
 E_{2 3}\,\raisebox{1.8ex}{$\ytableaushort{1123,2334,456,5}$}=\sqrt{\frac{5}{6}}\,\raisebox{1.8ex}{$\ytableaushort{1122,2334,456,5}$}+\sqrt{\frac{16}{6}}\,\raisebox{1.8ex}{$\ytableaushort{1123,2234,456,5}$}.
\end{align}
Finally, from the successive applications of the commutation relations Eq. \eqref{commutation} and from $E_{ij}=E_{ji}^{\dag}$, one gets the matrix representing the $\mathrm{SU}(N)$ FHM Hamiltonian $H$ [cf. Eq.~\ref{Hamiltonian}]
 in the irrep $\bar{\alpha}$, which corresponds to the $\mathrm{SU}(N)$  irrep $\alpha$.

 \begin{figure}[h!]
\includegraphics[width=0.4\textwidth]{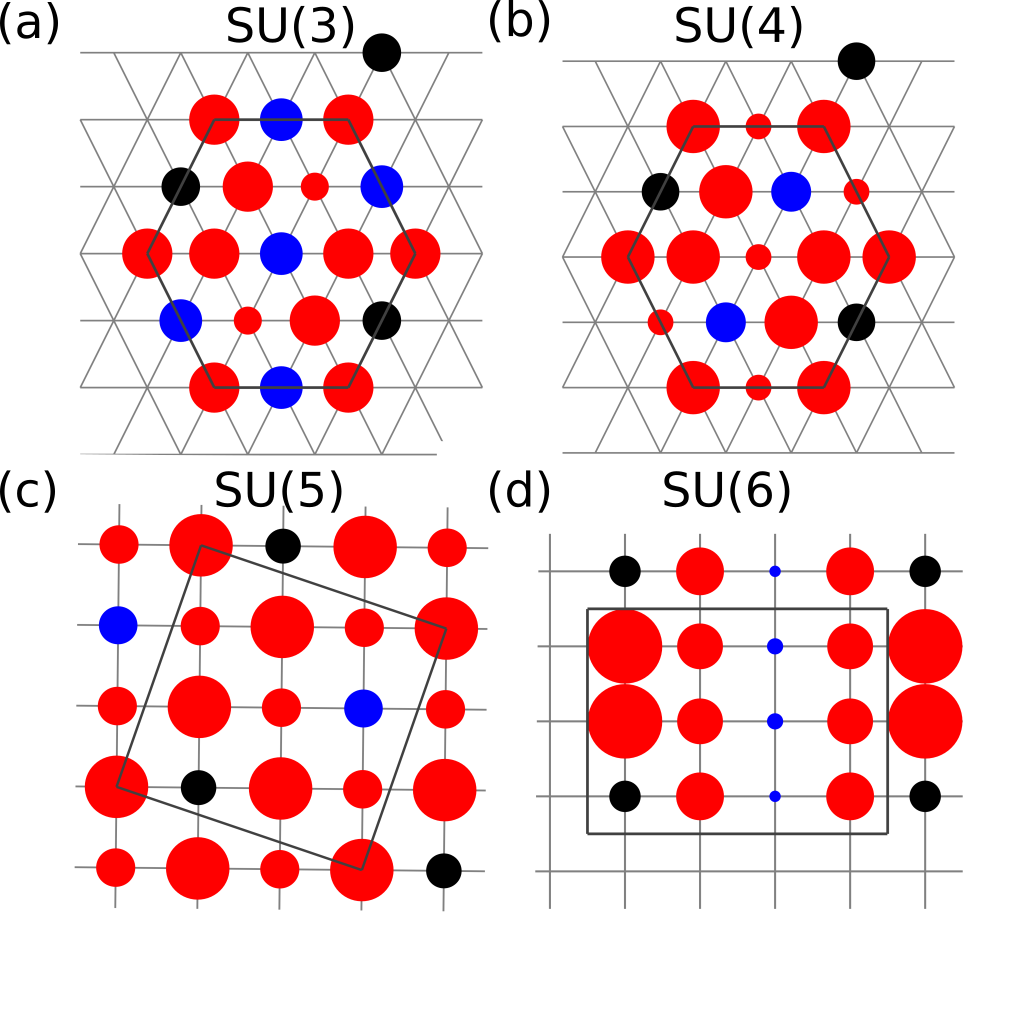}
\caption{Correlation patterns of the ground states of the FHM at filling $1/N$ for $U=10$ in (a),(c), d) and for  $U=12$ in (b). They are defined as $\langle P_{1j}\rangle -1/N$, where $P_{1j}\equiv -1+E_{1j}E_{j1}$, with the reference site  $1$ in black, and $j$ the site indices being blue (red) for positive (negative) correlation, with area proportional to its absolute value. At the top, triangular lattice with $L=12$ sites, 
 for $\mathrm{SU}(3)$ (a)) compatible with the three-sublattice N\'eel order~\cite{Bauer_2012,Tsunetsugu_2006,Laeuchli_2006}
and for $\mathrm{SU}(4)$ (in b)) compatible with the four-sublattice N\'eel order~\cite{Mila_2003}.
(c) $\mathrm{SU}(5)$ on the cluster $\sqrt{10}\times\sqrt{10}$, pattern compatible with the (chess) knight move LRO~\cite{nataf_exact_2014};
d) $\mathrm{SU}(6)$ on the $3 \times 4$ cluster, compatible with the $\mathrm{SU}(6)$ plaquette state~\cite{Gauthe_2021}.
}
\label{fig_correlations}
\end{figure}
 
 \begin{figure}[h!]
\includegraphics[width=0.5\textwidth]{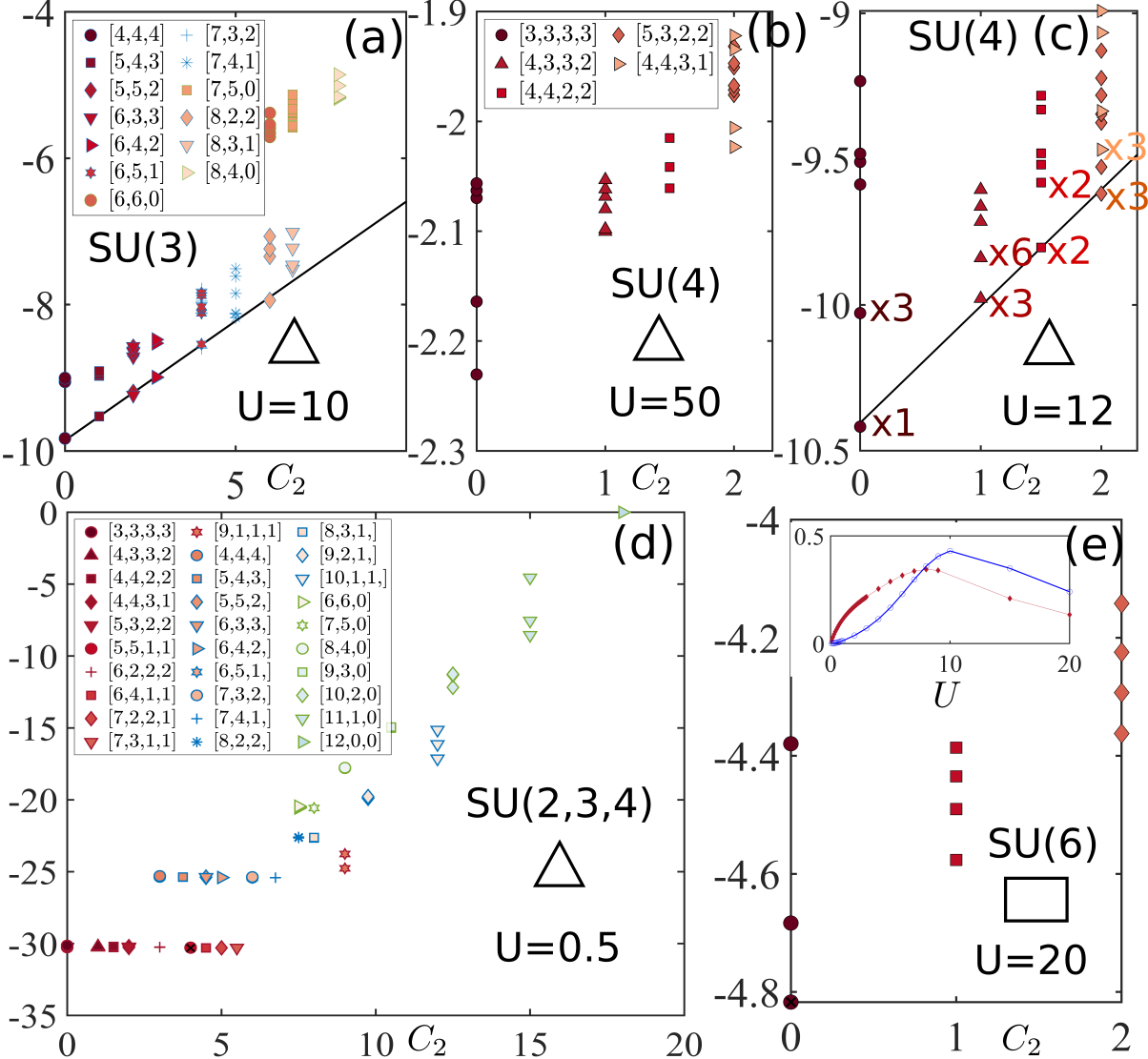}
\caption{Energy spectra of the FHM in Eq.~\eqref{Hamiltonian} (with $t_{ij}\equiv 1$ at filling $1/N$)  as a function of the quadratic Casimir $C_2$ for various values of $U$ and $N$ for the $L=12$ periodic triangular lattice (in a-d) and for the $L=4 \times 3$ periodic square lattice in (e), where we focused on $N=6$. Note that the constant $LU/2$ has been withdrawn.
(c) The Atos is reminiscent of the one revealing the four-sublattice order in the HM with nnn couplings~\cite{Mila_2003}.
Inset of (e): spin (singlet) gap in blue (red).
}
\label{fig_tos}
\end{figure}

 We have applied this theory to study the FHM of Eq. \eqref{Hamiltonian} for uniform nearest neighbors hopping $t_{ij}\equiv t=1$ as a function of $U$ at filling $1/N$, starting from the Heisenberg limit ($U\rightarrow \infty$) and diminishing $U$.
 The ground states of the  $\mathrm{SU}(3)$ HM on the triangular lattice (TL) and of the $\mathrm{SU}(5)$ HM on the square lattice (SL) are both N\'eel LRO states, with a three-sublattice ordering pattern  for $\mathrm{SU}(3)$~\cite{Bauer_2012,Tsunetsugu_2006,Laeuchli_2006} 
 and a (chess) knight move pattern for $\mathrm{SU}(5)$~\cite{nataf_exact_2014}. We give evidence of such orders by calculating the simple correlation patterns of the exact ground states of the FHM in Fig.~\ref{fig_correlations} for $U=10$.
 Moreover,  the energy spectra plotted as a function of the quadratic Casimir $C_2$~\cite{supp_mat,Paldus_2021} of the different irreps $\alpha$  exhibit an Anderson tower of states (Atos) 
which reveals the continuous symmetry breaking of $\mathrm{SU}(3)$ (respectively, $\mathrm{SU}(5)$) as shown in Fig.~\ref{fig_tos} (respectively, in the Supplemental Material~\cite{supp_mat}). 
We have checked the convergence  in the  limit $U\rightarrow \infty$ within each irrep $\alpha$ of the eigenenergies toward those of the HMs with the factor $2/U$. In fact, the group theory coefficients used in the protocol for the $\mathrm{SU}(N)$ FHM converge \cite{supp_mat} toward the ones needed in the algorithm for the $\mathrm{SU}(N)$ HM \cite{nataf_exact_2014}.

 While diminishing $U$, the structure of the energy spectra stays the same up to $U \sim 2.5$ (respectively, $U \sim 1$) for $\mathrm{SU}(3)$ on the $L=12$ TL (respectively, $\mathrm{SU}(5)$ on the $L=10$ SL). Then, some energy plateau as a function of $C_2$ appears for smaller $U$,
which is also true for $\mathrm{SU}(2)$ on the TL, as shown in Fig.~\ref{fig_tos}. 
Such a system should be in the metallic phase for  $U \lesssim 8.5 $, as expected from density-matrix renormalization group (DMRG) simulations on large cylinders~\cite{Szasz_2020}, so that the plateau could be a signature of the metallic phase in the weak coupling limit.
To further characterize the metallic phase and to locate its boundary, we show in Fig.~\ref{gaps} and in the Supplemental Material ~\cite{supp_mat} the charge gap defined by $\Delta_c=E_0(M=L+1)+E_0(M=L-1)-2E_0(M=L)$, where $E_0(M)$ is
the minimal energy for the lattice with $M$ fermions, which implies the diagonalization over all the relevant $M=L, L\pm 1$ box irreps $\bar{\alpha}$.
It suggests that the metallic phase develops for $U  \leq U_{c} = 9.8 (\pm 0.4)$ for $\mathrm{SU}(3)$ on a TL and for $U  \leq U_{c} = 8.75 (\pm 0.15)$ for $\mathrm{SU}(5)$ on a SL, with apparently no intermediate phase between the latter and the LRO in the large $U$ limit.
The scenario of successive LRO phases, with different antiferromagnetic orders, like what was numerically observed in the $\mathrm{SU}(3)$ FHM on the SL~\cite{Feng_2023}, does not seem to occur here, since the correlation patterns
are monotonic~\cite{supp_mat}. However, the sizes of the clusters within reach of our ED method do not exclude such a scenario in the bulk limit.

\begin{figure}[h!]
\includegraphics[width=0.5\textwidth]{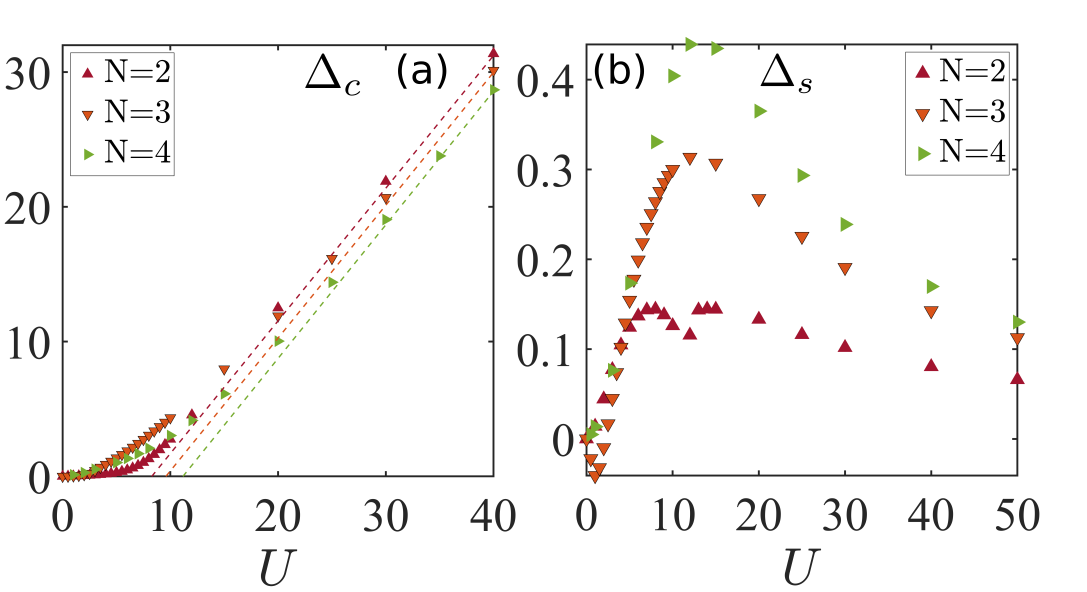}
\caption{(a) Charge gaps $\Delta_{\rm{c}}$ for the FHM on the $L=12$ sites triangular lattice for $N=2,3,$ and 4 at filling $1/N$. The dashed lines (cf. the Supplemental Material~\cite{supp_mat} for the fitting procedure) cross the $x$ axis at $U \simeq 8.6 $ for $N=2$, $U \simeq 9.8 $ for $N=3$,  and $U \simeq 11.2$ for $N=4$ separating the small $U$ metallic phase from the Mott insulators.
(b) $\mathrm{SU}(N)$ spin gaps $\Delta_{\rm{s}}$.}
\label{gaps}
\end{figure}

The presence of an in-between phase, detectable on finite-size clusters, might occur when the HM limit is not a LRO, such as for $\mathrm{SU}(3)$ on the honeycomb lattice~\cite{Corboz_2019}.
The  $\mathrm{SU}(4)$ FHM on the TL might enter into this category as the HM limit is a gapless quantum spin liquid~\cite{Mila_2003,Keselman_2020,Jin_2022}.
In fact, while there are four low-lying energy $\mathrm{SU}(4)$ singlet states for $U=50$ on the $L=12$ site TL, when decreasing the interaction, the singlet gap (defined as the gap within the singlet irrep $\alpha_{\mathcal{S},M}$) starts increasing
and an emerging Atos appears for $U \lesssim 15$ (cf Fig.~\ref{fig_tos} with both $U=12$ and $U=50$), which is similar to the one occurring in a pseudo-HM with next-nearest neighbour (nnn) couplings on the same $L=12$ TL cluster~\cite{Mila_2003}.
The nnn couplings, which are present at order 4 in $t/U$ in the large $U$ limit~\cite{Boos_2020}, were shown to stabilize a four-sublattice order~\cite{Mila_2003}, also apparent in the correlation patterns of our Fig.~\ref{fig_correlations}.
From the charge gap shown in Fig.~\ref{gaps}, the boundary between this phase and the metallic phase occurs around $U=U_c=11.2 \pm 0.2$. Like for the other systems, such a value slightly changes for larger $L$ due to finite-size effects, as illustrated for $\mathrm{SU}(2)$, $L=12$, and $L=16$ in the Supplemental Material~\cite{supp_mat}. We also show, in Fig.~\ref{gaps} (b), the $\mathrm{SU}(N)$ spin gaps $\Delta_{\rm{s}}$ defined as the difference between the minimal energy  of the $\mathrm{SU}(N)$ adjoint irrep sector (corresponding to $\alpha=[M/N+1,M/N, \dots, M/N-1]$)  and that of the singlet sector (i.e., for $\alpha=\alpha_{\mathcal{S},M}$). The spin gaps $\Delta_{\rm{s}}$, which are also impacted by finite-size effects~\cite{supp_mat}, exhibit some peaks at values of $U$ that roughly match with the values $U_c$ for each $N$.
 
Finally, we have investigated the $\mathrm{SU}(6)$ FHM on the SL, as the HM limit is also not a LRO but a plaquette state~\cite{Gauthe_2021}. Through the correlation pattern of the GS on a $4 \times 3$ periodic SL for $U=10$  in Fig.~\ref{fig_correlations},
we found some evidence of the six-sites plaquette state in the Mott phase, a feature confirmed by the presence of two low-lying energy $\mathrm{SU}(6)$ singlet states, compatible with the periodic boundary conditions,  for large $U$  ( i.e., $U=20$, cf. Fig.~\ref{fig_tos}).
When $U$ decreases, the spin gap becomes smaller than the singlet gap, suggesting a change of phases. However, with the current version of our code and with the limitation of our computational resources, the necessary calculation of the charge gap was too demanding,
leaving open both the question of the size of the metallic phase and the presence of some intermediate phase.

 To conclude, we found an efficient protocol to perform ED of the FHM on $L-$site clusters directly in each $\mathrm{SU}(N)$ irrep, which uses the set of ssYT (or GT patterns) as a convenient basis with matrix elements of the $\mathrm{U}(L)$ group generators.
 This approach, which generalizes the use of SYT for $\mathrm{SU}(N)$ in HMs~\cite{nataf_exact_2014},  dramatically reduces the dimension of the matrices to diagonalize. We applied our method to study the survival of the $\mathrm{SU}(N)$ Mott phases from  $N=3$ to $N=6$ on TL and SL when the on-site interaction $U$ decreases.
 In particular, we found an emerging intermediate LRO phase for $\mathrm{SU}(4)$ on the TL, reminiscent of the four-sublattice order in the HM with nnn couplings~\cite{Mila_2003}.
  
  Among the perspectives, since the $\mathrm{SU}(N)$ FHM can be seen as a fine-tuned version of the $\mathrm{Sp}(N)$ FHM \cite{Wu_2005},
 one could generalize our approach to Hamiltonians invariant under $\mathrm{Sp}(N)$ \cite{Wu_review_2006,Coleman_2008,Ramires_2017}.
 It would be numerically helpful, as the dimension of the $\mathrm{Sp}(N)$ singlet sector is also much smaller than that of the sector used in traditional ED \cite{supp_mat}.
 However, GT-type basis for finite dimensional irreps of $\mathrm{Sp}(N)$ are more complicated to handle than those for the irreps of  $\mathrm{SU}(N)$ \cite{Molev_1999,Molev_2002}.
Other perspectives would be the implementation  of the ssYT basis in tensor networks and DMRG algorithms, 
in a fashion similar to what was done with the SYT for the HM~\cite{nataf_density_2018,gozel_2020}, and the combination of both the implementation of  the $\mathrm{SU}(N)$ and of the spatial symmetries in ED.

 \vspace{0.5cm}

We acknowledge F. Mila for fruitful discussions and S. Gozel for a critical reading of the manuscript, as well as A. Laeuchli and C. Ganahl who gave us some 15 digits $\mathrm{SU}(3)$
FHM eigenenergies \cite{Boos_2020} for data comparisons.
 This work has been supported by an Emergence grant from CNRS Physique.

{\it Appendix.}---The matrix elements of the infinitesimal generators between equal or consecutive sites $E_{p p}$, $E_{p-1 p}$, $E_{p p-1}$, take a simple form on the basis of the ssYT. Calling $\vert \nu \rangle$ a ssYT, one has  for $p=1,\dots,L$,
\begin{align}
E_{p  p} \vert \nu \rangle &=\text{Cardinal}\big{\{} p \in \nu \big{\}}\vert \nu \rangle, 
\end{align}
where $\text{Cardinal}\big{\{} p \in \nu \big{\}}$ is equal to the number of occurrences of $p$ inside $\vert \nu \rangle$, corresponding to the occupation number on site $p$ (cf. Fig.~\ref{fig1} for some examples).

Second, calling $\vert \nu \rangle$ a ssYT, one has  for $p=2,\dots,L$,
\begin{equation}
E_{p-1 p} \vert \nu \rangle = \sum_{j=1}^{p-1} a^j_{p-1} F^j_{p-1} \vert \nu \rangle,
\end{equation}
where the tableau operators $F^j_{p-1} $ transform the number $p$ in the $j$th  row in $\vert \nu \rangle$ into $p-1$.
As for the coefficients  $ a^j_{p-1} $, which vanish in the case where such a transformation is not possible, either because there is no $p$ in the $j$th row of $\vert \nu \rangle$ or because the resulting tableau is not a proper ssYT, they read  \cite{Vilenkin_vol3,supp_mat}
\begin{equation}
a^j_{p-1}=  \left | \frac{\prod_{i=1}^p (l_{i,p}-l_{j,p-1})\prod_{i=1}^{p-2} (l_{i,p-2}-l_{j,p-1}-1)}{ \prod_{i\neq j} (l_{i,p-1}-l_{j,p-1})\prod_{i\neq j} (l_{i,p-1}-l_{j,p-1}-1)} \right | ^{1/2},
\end{equation}
where $l_{k,q}=m_{k,q}-k$ with $m_{k,q}$ as the length of the $k$th row of the subtableau that remains when we delete all the boxes containing numbers $>q$ in $\vert \nu \rangle$.
Finally,  from $E_{ij}=E_{ji}^{\dag}$, we obtain the matrix elements of $E_{p p-1} $ for $p=2,\dots,L$.

\bibliographystyle{apsrev4-1}
\bibliography{main.bib}

\begin{thebibliography}{87}%
\makeatletter
\providecommand \@ifxundefined [1]{%
 \@ifx{#1\undefined}
}%
\providecommand \@ifnum [1]{%
 \ifnum #1\expandafter \@firstoftwo
 \else \expandafter \@secondoftwo
 \fi
}%
\providecommand \@ifx [1]{%
 \ifx #1\expandafter \@firstoftwo
 \else \expandafter \@secondoftwo
 \fi
}%
\providecommand \natexlab [1]{#1}%
\providecommand \enquote  [1]{``#1''}%
\providecommand \bibnamefont  [1]{#1}%
\providecommand \bibfnamefont [1]{#1}%
\providecommand \citenamefont [1]{#1}%
\providecommand \href@noop [0]{\@secondoftwo}%
\providecommand \href [0]{\begingroup \@sanitize@url \@href}%
\providecommand \@href[1]{\@@startlink{#1}\@@href}%
\providecommand \@@href[1]{\endgroup#1\@@endlink}%
\providecommand \@sanitize@url [0]{\catcode `\\12\catcode `\$12\catcode `\&12\catcode `\#12\catcode `\^12\catcode `\_12\catcode `\%12\relax}%
\providecommand \@@startlink[1]{}%
\providecommand \@@endlink[0]{}%
\providecommand \url  [0]{\begingroup\@sanitize@url \@url }%
\providecommand \@url [1]{\endgroup\@href {#1}{\urlprefix }}%
\providecommand \urlprefix  [0]{URL }%
\providecommand \Eprint [0]{\href }%
\providecommand \doibase [0]{http://dx.doi.org/}%
\providecommand \selectlanguage [0]{\@gobble}%
\providecommand \bibinfo  [0]{\@secondoftwo}%
\providecommand \bibfield  [0]{\@secondoftwo}%
\providecommand \translation [1]{[#1]}%
\providecommand \BibitemOpen [0]{}%
\providecommand \bibitemStop [0]{}%
\providecommand \bibitemNoStop [0]{.\EOS\space}%
\providecommand \EOS [0]{\spacefactor3000\relax}%
\providecommand \BibitemShut  [1]{\csname bibitem#1\endcsname}%
\let\auto@bib@innerbib\@empty
\bibitem [{\citenamefont {Hubbard}(1963)}]{Hubbard_1963}%
  \BibitemOpen
  \bibfield  {author} {\bibinfo {author} {\bibfnamefont {J.}~\bibnamefont {Hubbard}},\ }\href {\doibase 10.1098/rspa.1963.0204} {\bibfield  {journal} {\bibinfo  {journal} {Proc. R. Soc. A}\ }\textbf {\bibinfo {volume} {276}},\ \bibinfo {pages} {238} (\bibinfo {year} {1963})}\BibitemShut {NoStop}%
\bibitem [{\citenamefont {Gutzwiller}(1963)}]{Gutzwiller_1963}%
  \BibitemOpen
  \bibfield  {author} {\bibinfo {author} {\bibfnamefont {M.~C.}\ \bibnamefont {Gutzwiller}},\ }\href {\doibase 10.1103/PhysRevLett.10.159} {\bibfield  {journal} {\bibinfo  {journal} {Phys. Rev. Lett.}\ }\textbf {\bibinfo {volume} {10}},\ \bibinfo {pages} {159} (\bibinfo {year} {1963})}\BibitemShut {NoStop}%
\bibitem [{\citenamefont {Scalapino}(2012)}]{Scalapino_2012}%
  \BibitemOpen
  \bibfield  {author} {\bibinfo {author} {\bibfnamefont {D.~J.}\ \bibnamefont {Scalapino}},\ }\href {\doibase 10.1103/RevModPhys.84.1383} {\bibfield  {journal} {\bibinfo  {journal} {Rev. Mod. Phys.}\ }\textbf {\bibinfo {volume} {84}},\ \bibinfo {pages} {1383} (\bibinfo {year} {2012})}\BibitemShut {NoStop}%
\bibitem [{\citenamefont {Anderson}(1987)}]{Anderson_1987}%
  \BibitemOpen
  \bibfield  {author} {\bibinfo {author} {\bibfnamefont {P.~W.}\ \bibnamefont {Anderson}},\ }\href {\doibase 10.1126/science.235.4793.1196} {\bibfield  {journal} {\bibinfo  {journal} {Science}\ }\textbf {\bibinfo {volume} {235}},\ \bibinfo {pages} {1196} (\bibinfo {year} {1987})}\BibitemShut {NoStop}%
\bibitem [{\citenamefont {Zhang}\ and\ \citenamefont {Rice}(1988)}]{Rice_1988}%
  \BibitemOpen
  \bibfield  {author} {\bibinfo {author} {\bibfnamefont {F.~C.}\ \bibnamefont {Zhang}}\ and\ \bibinfo {author} {\bibfnamefont {T.~M.}\ \bibnamefont {Rice}},\ }\href {\doibase 10.1103/PhysRevB.37.3759} {\bibfield  {journal} {\bibinfo  {journal} {Phys. Rev. B}\ }\textbf {\bibinfo {volume} {37}},\ \bibinfo {pages} {3759} (\bibinfo {year} {1988})}\BibitemShut {NoStop}%
\bibitem [{\citenamefont {Arovas}\ \emph {et~al.}(2022)\citenamefont {Arovas}, \citenamefont {Berg}, \citenamefont {Kivelson},\ and\ \citenamefont {Raghu}}]{review_Arovas_2022}%
  \BibitemOpen
  \bibfield  {author} {\bibinfo {author} {\bibfnamefont {D.~P.}\ \bibnamefont {Arovas}}, \bibinfo {author} {\bibfnamefont {E.}~\bibnamefont {Berg}}, \bibinfo {author} {\bibfnamefont {S.~A.}\ \bibnamefont {Kivelson}}, \ and\ \bibinfo {author} {\bibfnamefont {S.}~\bibnamefont {Raghu}},\ }\href {\doibase 10.1146/annurev-conmatphys-031620-102024} {\bibfield  {journal} {\bibinfo  {journal} {Annu. Rev. Condens. Matter Phys.}\ }\textbf {\bibinfo {volume} {13}},\ \bibinfo {pages} {239} (\bibinfo {year} {2022})}\BibitemShut {NoStop}%
\bibitem [{\citenamefont {Qin}\ \emph {et~al.}(2022)\citenamefont {Qin}, \citenamefont {Sch\"{a}fer}, \citenamefont {Andergassen}, \citenamefont {Corboz},\ and\ \citenamefont {Gull}}]{review_Corboz_2022}%
  \BibitemOpen
  \bibfield  {author} {\bibinfo {author} {\bibfnamefont {M.}~\bibnamefont {Qin}}, \bibinfo {author} {\bibfnamefont {T.}~\bibnamefont {Sch\"{a}fer}}, \bibinfo {author} {\bibfnamefont {S.}~\bibnamefont {Andergassen}}, \bibinfo {author} {\bibfnamefont {P.}~\bibnamefont {Corboz}}, \ and\ \bibinfo {author} {\bibfnamefont {E.}~\bibnamefont {Gull}},\ }\href {\doibase 10.1146/annurev-conmatphys-090921-033948} {\bibfield  {journal} {\bibinfo  {journal} {Annu. Rev. Condens. Matter Phys.}\ }\textbf {\bibinfo {volume} {13}},\ \bibinfo {pages} {275} (\bibinfo {year} {2022})}\BibitemShut {NoStop}%
\bibitem [{\citenamefont {Assaraf}\ \emph {et~al.}(1999)\citenamefont {Assaraf}, \citenamefont {Azaria}, \citenamefont {Caffarel},\ and\ \citenamefont {Lecheminant}}]{Assaraf_1999}%
  \BibitemOpen
  \bibfield  {author} {\bibinfo {author} {\bibfnamefont {R.}~\bibnamefont {Assaraf}}, \bibinfo {author} {\bibfnamefont {P.}~\bibnamefont {Azaria}}, \bibinfo {author} {\bibfnamefont {M.}~\bibnamefont {Caffarel}}, \ and\ \bibinfo {author} {\bibfnamefont {P.}~\bibnamefont {Lecheminant}},\ }\href {\doibase 10.1103/PhysRevB.60.2299} {\bibfield  {journal} {\bibinfo  {journal} {Phys. Rev. B}\ }\textbf {\bibinfo {volume} {60}},\ \bibinfo {pages} {2299} (\bibinfo {year} {1999})}\BibitemShut {NoStop}%
\bibitem [{\citenamefont {Honerkamp}\ and\ \citenamefont {Hofstetter}(2004)}]{honerkamp_ultrcold_2004}%
  \BibitemOpen
  \bibfield  {author} {\bibinfo {author} {\bibfnamefont {C.}~\bibnamefont {Honerkamp}}\ and\ \bibinfo {author} {\bibfnamefont {W.}~\bibnamefont {Hofstetter}},\ }\href {https://link.aps.org/doi/10.1103/PhysRevLett.92.170403} {\bibfield  {journal} {\bibinfo  {journal} {Phys. Rev. Lett.}\ }\textbf {\bibinfo {volume} {92}},\ \bibinfo {pages} {170403} (\bibinfo {year} {2004})}\BibitemShut {NoStop}%
\bibitem [{\citenamefont {Capponi}\ \emph {et~al.}(2016)\citenamefont {Capponi}, \citenamefont {Lecheminant},\ and\ \citenamefont {Totsuka}}]{capponi_phases_2016}%
  \BibitemOpen
  \bibfield  {author} {\bibinfo {author} {\bibfnamefont {S.}~\bibnamefont {Capponi}}, \bibinfo {author} {\bibfnamefont {P.}~\bibnamefont {Lecheminant}}, \ and\ \bibinfo {author} {\bibfnamefont {K.}~\bibnamefont {Totsuka}},\ }\href {http://www.sciencedirect.com/science/article/pii/S0003491616000130} {\bibfield  {journal} {\bibinfo  {journal} {Ann. Phys. (Amsterdam)}\ }\textbf {\bibinfo {volume} {367}},\ \bibinfo {pages} {50 } (\bibinfo {year} {2016})}\BibitemShut {NoStop}%
\bibitem [{\citenamefont {Affleck}(1986)}]{affleck_exact_1986}%
  \BibitemOpen
  \bibfield  {author} {\bibinfo {author} {\bibfnamefont {I.}~\bibnamefont {Affleck}},\ }\href {http://www.sciencedirect.com/science/article/pii/0550321386901677} {\bibfield  {journal} {\bibinfo  {journal} {Nucl. Phys.}\ }\textbf {\bibinfo {volume} {B265}},\ \bibinfo {pages} {409 } (\bibinfo {year} {1986})}\BibitemShut {NoStop}%
\bibitem [{\citenamefont {Affleck}\ and\ \citenamefont {Marston}(1988)}]{Affleck_1988}%
  \BibitemOpen
  \bibfield  {author} {\bibinfo {author} {\bibfnamefont {I.}~\bibnamefont {Affleck}}\ and\ \bibinfo {author} {\bibfnamefont {J.~B.}\ \bibnamefont {Marston}},\ }\href {\doibase 10.1103/PhysRevB.37.3774} {\bibfield  {journal} {\bibinfo  {journal} {Phys. Rev. B}\ }\textbf {\bibinfo {volume} {37}},\ \bibinfo {pages} {3774} (\bibinfo {year} {1988})}\BibitemShut {NoStop}%
\bibitem [{\citenamefont {Rokhsar}(1990)}]{Rokhsar_1990}%
  \BibitemOpen
  \bibfield  {author} {\bibinfo {author} {\bibfnamefont {D.~S.}\ \bibnamefont {Rokhsar}},\ }\href {\doibase 10.1103/PhysRevB.42.2526} {\bibfield  {journal} {\bibinfo  {journal} {Phys. Rev. B}\ }\textbf {\bibinfo {volume} {42}},\ \bibinfo {pages} {2526} (\bibinfo {year} {1990})}\BibitemShut {NoStop}%
\bibitem [{\citenamefont {Marder}\ \emph {et~al.}(1990)\citenamefont {Marder}, \citenamefont {Papanicolaou},\ and\ \citenamefont {Psaltakis}}]{Marder_1990}%
  \BibitemOpen
  \bibfield  {author} {\bibinfo {author} {\bibfnamefont {M.}~\bibnamefont {Marder}}, \bibinfo {author} {\bibfnamefont {N.}~\bibnamefont {Papanicolaou}}, \ and\ \bibinfo {author} {\bibfnamefont {G.~C.}\ \bibnamefont {Psaltakis}},\ }\href {\doibase 10.1103/PhysRevB.41.6920} {\bibfield  {journal} {\bibinfo  {journal} {Phys. Rev. B}\ }\textbf {\bibinfo {volume} {41}},\ \bibinfo {pages} {6920} (\bibinfo {year} {1990})}\BibitemShut {NoStop}%
\bibitem [{\citenamefont {Pati}\ \emph {et~al.}(1998)\citenamefont {Pati}, \citenamefont {Singh},\ and\ \citenamefont {Khomskii}}]{Khomskii_1998}%
  \BibitemOpen
  \bibfield  {author} {\bibinfo {author} {\bibfnamefont {S.~K.}\ \bibnamefont {Pati}}, \bibinfo {author} {\bibfnamefont {R.~R.~P.}\ \bibnamefont {Singh}}, \ and\ \bibinfo {author} {\bibfnamefont {D.~I.}\ \bibnamefont {Khomskii}},\ }\href {\doibase 10.1103/PhysRevLett.81.5406} {\bibfield  {journal} {\bibinfo  {journal} {Phys. Rev. Lett.}\ }\textbf {\bibinfo {volume} {81}},\ \bibinfo {pages} {5406} (\bibinfo {year} {1998})}\BibitemShut {NoStop}%
\bibitem [{\citenamefont {Yamada}\ \emph {et~al.}(2018)\citenamefont {Yamada}, \citenamefont {Oshikawa},\ and\ \citenamefont {Jackeli}}]{Yamada_2018}%
  \BibitemOpen
  \bibfield  {author} {\bibinfo {author} {\bibfnamefont {M.~G.}\ \bibnamefont {Yamada}}, \bibinfo {author} {\bibfnamefont {M.}~\bibnamefont {Oshikawa}}, \ and\ \bibinfo {author} {\bibfnamefont {G.}~\bibnamefont {Jackeli}},\ }\href {\doibase 10.1103/PhysRevLett.121.097201} {\bibfield  {journal} {\bibinfo  {journal} {Phys. Rev. Lett.}\ }\textbf {\bibinfo {volume} {121}},\ \bibinfo {pages} {097201} (\bibinfo {year} {2018})}\BibitemShut {NoStop}%
\bibitem [{\citenamefont {Zhang}\ \emph {et~al.}(2021)\citenamefont {Zhang}, \citenamefont {Sheng},\ and\ \citenamefont {Vishwanath}}]{zhang2021}%
  \BibitemOpen
  \bibfield  {author} {\bibinfo {author} {\bibfnamefont {Y.-H.}\ \bibnamefont {Zhang}}, \bibinfo {author} {\bibfnamefont {D.~N.}\ \bibnamefont {Sheng}}, \ and\ \bibinfo {author} {\bibfnamefont {A.}~\bibnamefont {Vishwanath}},\ }\href {\doibase 10.1103/PhysRevLett.127.247701} {\bibfield  {journal} {\bibinfo  {journal} {Phys. Rev. Lett.}\ }\textbf {\bibinfo {volume} {127}},\ \bibinfo {pages} {247701} (\bibinfo {year} {2021})}\BibitemShut {NoStop}%
\bibitem [{\citenamefont {Wu}\ \emph {et~al.}(2003)\citenamefont {Wu}, \citenamefont {Hu},\ and\ \citenamefont {Zhang}}]{wu_exact_2003}%
  \BibitemOpen
  \bibfield  {author} {\bibinfo {author} {\bibfnamefont {C.}~\bibnamefont {Wu}}, \bibinfo {author} {\bibfnamefont {J.-P.}\ \bibnamefont {Hu}}, \ and\ \bibinfo {author} {\bibfnamefont {S.-c.}\ \bibnamefont {Zhang}},\ }\href {https://link.aps.org/doi/10.1103/PhysRevLett.91.186402} {\bibfield  {journal} {\bibinfo  {journal} {Phys. Rev. Lett.}\ }\textbf {\bibinfo {volume} {91}},\ \bibinfo {pages} {186402} (\bibinfo {year} {2003})}\BibitemShut {NoStop}%
\bibitem [{\citenamefont {Wu}(2006)}]{Wu_review_2006}%
  \BibitemOpen
  \bibfield  {author} {\bibinfo {author} {\bibfnamefont {C.}~\bibnamefont {Wu}},\ }\href {\doibase 10.1142/S0217984906012213} {\bibfield  {journal} {\bibinfo  {journal} {Mod. Phys. Lett. B}\ }\textbf {\bibinfo {volume} {20}},\ \bibinfo {pages} {1707} (\bibinfo {year} {2006})}\BibitemShut {NoStop}%
\bibitem [{\citenamefont {Gorshkov}\ \emph {et~al.}(2010)\citenamefont {Gorshkov}, \citenamefont {Hermele}, \citenamefont {Gurarie}, \citenamefont {Xu}, \citenamefont {Julienne}, \citenamefont {Ye}, \citenamefont {Zoller}, \citenamefont {Demler}, \citenamefont {Lukin},\ and\ \citenamefont {Rey}}]{gorshkov_two_2010}%
  \BibitemOpen
  \bibfield  {author} {\bibinfo {author} {\bibfnamefont {A.~V.}\ \bibnamefont {Gorshkov}}, \bibinfo {author} {\bibfnamefont {M.}~\bibnamefont {Hermele}}, \bibinfo {author} {\bibfnamefont {V.}~\bibnamefont {Gurarie}}, \bibinfo {author} {\bibfnamefont {C.}~\bibnamefont {Xu}}, \bibinfo {author} {\bibfnamefont {P.~S.}\ \bibnamefont {Julienne}}, \bibinfo {author} {\bibfnamefont {J.}~\bibnamefont {Ye}}, \bibinfo {author} {\bibfnamefont {P.}~\bibnamefont {Zoller}}, \bibinfo {author} {\bibfnamefont {E.}~\bibnamefont {Demler}}, \bibinfo {author} {\bibfnamefont {M.~D.}\ \bibnamefont {Lukin}}, \ and\ \bibinfo {author} {\bibfnamefont {A.}~\bibnamefont {Rey}},\ }\href {https://www.nature.com/articles/nphys1535} {\bibfield  {journal} {\bibinfo  {journal} {Nat. Phys.}\ }\textbf {\bibinfo {volume} {6}},\ \bibinfo {pages} {289} (\bibinfo {year} {2010})}\BibitemShut {NoStop}%
\bibitem [{\citenamefont {Cazalilla}\ and\ \citenamefont {Rey}(2014)}]{Cazalilla_2014}%
  \BibitemOpen
  \bibfield  {author} {\bibinfo {author} {\bibfnamefont {M.~A.}\ \bibnamefont {Cazalilla}}\ and\ \bibinfo {author} {\bibfnamefont {A.~M.}\ \bibnamefont {Rey}},\ }\href {\doibase 10.1088/0034-4885/77/12/124401} {\bibfield  {journal} {\bibinfo  {journal} {Rep. Prog. Phys.}\ }\textbf {\bibinfo {volume} {77}},\ \bibinfo {pages} {124401} (\bibinfo {year} {2014})}\BibitemShut {NoStop}%
\bibitem [{\citenamefont {Taie}\ \emph {et~al.}(2012)\citenamefont {Taie}, \citenamefont {Yamazaki}, \citenamefont {Sugawa},\ and\ \citenamefont {Takahashi}}]{taie_su6_2012}%
  \BibitemOpen
  \bibfield  {author} {\bibinfo {author} {\bibfnamefont {S.}~\bibnamefont {Taie}}, \bibinfo {author} {\bibfnamefont {R.}~\bibnamefont {Yamazaki}}, \bibinfo {author} {\bibfnamefont {S.}~\bibnamefont {Sugawa}}, \ and\ \bibinfo {author} {\bibfnamefont {Y.}~\bibnamefont {Takahashi}},\ }\href {https://www.nature.com/articles/nphys2430} {\bibfield  {journal} {\bibinfo  {journal} {Nat. Phys.}\ }\textbf {\bibinfo {volume} {8}},\ \bibinfo {pages} {825} (\bibinfo {year} {2012})}\BibitemShut {NoStop}%
\bibitem [{\citenamefont {Hofrichter}\ \emph {et~al.}(2016)\citenamefont {Hofrichter}, \citenamefont {Riegger}, \citenamefont {Scazza}, \citenamefont {H\"ofer}, \citenamefont {Fernandes}, \citenamefont {Bloch},\ and\ \citenamefont {F\"olling}}]{hofrichter_direct_2016}%
  \BibitemOpen
  \bibfield  {author} {\bibinfo {author} {\bibfnamefont {C.}~\bibnamefont {Hofrichter}}, \bibinfo {author} {\bibfnamefont {L.}~\bibnamefont {Riegger}}, \bibinfo {author} {\bibfnamefont {F.}~\bibnamefont {Scazza}}, \bibinfo {author} {\bibfnamefont {M.}~\bibnamefont {H\"ofer}}, \bibinfo {author} {\bibfnamefont {D.~R.}\ \bibnamefont {Fernandes}}, \bibinfo {author} {\bibfnamefont {I.}~\bibnamefont {Bloch}}, \ and\ \bibinfo {author} {\bibfnamefont {S.}~\bibnamefont {F\"olling}},\ }\href {\doibase 10.1103/PhysRevX.6.021030} {\bibfield  {journal} {\bibinfo  {journal} {Phys. Rev. X}\ }\textbf {\bibinfo {volume} {6}},\ \bibinfo {pages} {021030} (\bibinfo {year} {2016})}\BibitemShut {NoStop}%
\bibitem [{\citenamefont {Abeln}\ \emph {et~al.}(2021)\citenamefont {Abeln}, \citenamefont {Sponselee}, \citenamefont {Diem}, \citenamefont {Pintul}, \citenamefont {Sengstock},\ and\ \citenamefont {Becker}}]{Becker_2021}%
  \BibitemOpen
  \bibfield  {author} {\bibinfo {author} {\bibfnamefont {B.}~\bibnamefont {Abeln}}, \bibinfo {author} {\bibfnamefont {K.}~\bibnamefont {Sponselee}}, \bibinfo {author} {\bibfnamefont {M.}~\bibnamefont {Diem}}, \bibinfo {author} {\bibfnamefont {N.}~\bibnamefont {Pintul}}, \bibinfo {author} {\bibfnamefont {K.}~\bibnamefont {Sengstock}}, \ and\ \bibinfo {author} {\bibfnamefont {C.}~\bibnamefont {Becker}},\ }\href {\doibase 10.1103/PhysRevA.103.033315} {\bibfield  {journal} {\bibinfo  {journal} {Phys. Rev. A}\ }\textbf {\bibinfo {volume} {103}},\ \bibinfo {pages} {033315} (\bibinfo {year} {2021})}\BibitemShut {NoStop}%
\bibitem [{\citenamefont {Taie}\ \emph {et~al.}(2022)\citenamefont {Taie}, \citenamefont {Ibarra-Garc{\'\i}a-Padilla}, \citenamefont {Nishizawa}, \citenamefont {Takasu}, \citenamefont {Kuno}, \citenamefont {Wei}, \citenamefont {Scalettar}, \citenamefont {Hazzard},\ and\ \citenamefont {Takahashi}}]{taie2020observation}%
  \BibitemOpen
  \bibfield  {author} {\bibinfo {author} {\bibfnamefont {S.}~\bibnamefont {Taie}}, \bibinfo {author} {\bibfnamefont {E.}~\bibnamefont {Ibarra-Garc{\'\i}a-Padilla}}, \bibinfo {author} {\bibfnamefont {N.}~\bibnamefont {Nishizawa}}, \bibinfo {author} {\bibfnamefont {Y.}~\bibnamefont {Takasu}}, \bibinfo {author} {\bibfnamefont {Y.}~\bibnamefont {Kuno}}, \bibinfo {author} {\bibfnamefont {H.-T.}\ \bibnamefont {Wei}}, \bibinfo {author} {\bibfnamefont {R.~T.}\ \bibnamefont {Scalettar}}, \bibinfo {author} {\bibfnamefont {K.~R.~A.}\ \bibnamefont {Hazzard}}, \ and\ \bibinfo {author} {\bibfnamefont {Y.}~\bibnamefont {Takahashi}},\ }\href {\doibase 10.1038/s41567-022-01725-6} {\bibfield  {journal} {\bibinfo  {journal} {Nat. Phys.}\ }\textbf {\bibinfo {volume} {18}},\ \bibinfo {pages} {1356} (\bibinfo {year} {2022})}\BibitemShut {NoStop}%
\bibitem [{\citenamefont {Tusi}\ \emph {et~al.}(2022)\citenamefont {Tusi}, \citenamefont {Franchi}, \citenamefont {Livi}, \citenamefont {Baumann}, \citenamefont {Benedicto~Orenes}, \citenamefont {Del~Re}, \citenamefont {Barfknecht}, \citenamefont {Zhou}, \citenamefont {Inguscio}, \citenamefont {Cappellini}, \citenamefont {Capone}, \citenamefont {Catani},\ and\ \citenamefont {Fallani}}]{Fallani_2022}%
  \BibitemOpen
  \bibfield  {author} {\bibinfo {author} {\bibfnamefont {D.}~\bibnamefont {Tusi}}, \bibinfo {author} {\bibfnamefont {L.}~\bibnamefont {Franchi}}, \bibinfo {author} {\bibfnamefont {L.~F.}\ \bibnamefont {Livi}}, \bibinfo {author} {\bibfnamefont {K.}~\bibnamefont {Baumann}}, \bibinfo {author} {\bibfnamefont {D.}~\bibnamefont {Benedicto~Orenes}}, \bibinfo {author} {\bibfnamefont {L.}~\bibnamefont {Del~Re}}, \bibinfo {author} {\bibfnamefont {R.~E.}\ \bibnamefont {Barfknecht}}, \bibinfo {author} {\bibfnamefont {T.~W.}\ \bibnamefont {Zhou}}, \bibinfo {author} {\bibfnamefont {M.}~\bibnamefont {Inguscio}}, \bibinfo {author} {\bibfnamefont {G.}~\bibnamefont {Cappellini}}, \bibinfo {author} {\bibfnamefont {M.}~\bibnamefont {Capone}}, \bibinfo {author} {\bibfnamefont {J.}~\bibnamefont {Catani}}, \ and\ \bibinfo {author} {\bibfnamefont {L.}~\bibnamefont {Fallani}},\ }\href {\doibase 10.1038/s41567-022-01726-5} {\bibfield  {journal} {\bibinfo  {journal} {Nat. Phys.}\ }\textbf {\bibinfo {volume} {18}},\ \bibinfo {pages}
  {1201} (\bibinfo {year} {2022})}\BibitemShut {NoStop}%
\bibitem [{\citenamefont {Pasqualetti}\ \emph {et~al.}(2024)\citenamefont {Pasqualetti}, \citenamefont {Bettermann}, \citenamefont {Darkwah~Oppong}, \citenamefont {Ibarra-Garc\'{\i}a-Padilla}, \citenamefont {Dasgupta}, \citenamefont {Scalettar}, \citenamefont {Hazzard}, \citenamefont {Bloch},\ and\ \citenamefont {F\"olling}}]{pasqualetti2023equation}%
  \BibitemOpen
  \bibfield  {author} {\bibinfo {author} {\bibfnamefont {G.}~\bibnamefont {Pasqualetti}}, \bibinfo {author} {\bibfnamefont {O.}~\bibnamefont {Bettermann}}, \bibinfo {author} {\bibfnamefont {N.}~\bibnamefont {Darkwah~Oppong}}, \bibinfo {author} {\bibfnamefont {E.}~\bibnamefont {Ibarra-Garc\'{\i}a-Padilla}}, \bibinfo {author} {\bibfnamefont {S.}~\bibnamefont {Dasgupta}}, \bibinfo {author} {\bibfnamefont {R.~T.}\ \bibnamefont {Scalettar}}, \bibinfo {author} {\bibfnamefont {K.~R.~A.}\ \bibnamefont {Hazzard}}, \bibinfo {author} {\bibfnamefont {I.}~\bibnamefont {Bloch}}, \ and\ \bibinfo {author} {\bibfnamefont {S.}~\bibnamefont {F\"olling}},\ }\href {\doibase 10.1103/PhysRevLett.132.083401} {\bibfield  {journal} {\bibinfo  {journal} {Phys. Rev. Lett.}\ }\textbf {\bibinfo {volume} {132}},\ \bibinfo {pages} {083401} (\bibinfo {year} {2024})}\BibitemShut {NoStop}%
\bibitem [{\citenamefont {Wang}\ \emph {et~al.}(2014)\citenamefont {Wang}, \citenamefont {Li}, \citenamefont {Cai}, \citenamefont {Zhou}, \citenamefont {Wang},\ and\ \citenamefont {Wu}}]{Wang_2014}%
  \BibitemOpen
  \bibfield  {author} {\bibinfo {author} {\bibfnamefont {D.}~\bibnamefont {Wang}}, \bibinfo {author} {\bibfnamefont {Y.}~\bibnamefont {Li}}, \bibinfo {author} {\bibfnamefont {Z.}~\bibnamefont {Cai}}, \bibinfo {author} {\bibfnamefont {Z.}~\bibnamefont {Zhou}}, \bibinfo {author} {\bibfnamefont {Y.}~\bibnamefont {Wang}}, \ and\ \bibinfo {author} {\bibfnamefont {C.}~\bibnamefont {Wu}},\ }\href {\doibase 10.1103/PhysRevLett.112.156403} {\bibfield  {journal} {\bibinfo  {journal} {Phys. Rev. Lett.}\ }\textbf {\bibinfo {volume} {112}},\ \bibinfo {pages} {156403} (\bibinfo {year} {2014})}\BibitemShut {NoStop}%
\bibitem [{\citenamefont {Xu}\ \emph {et~al.}(2018)\citenamefont {Xu}, \citenamefont {Barreiro}, \citenamefont {Wang},\ and\ \citenamefont {Wu}}]{Xu_2018}%
  \BibitemOpen
  \bibfield  {author} {\bibinfo {author} {\bibfnamefont {S.}~\bibnamefont {Xu}}, \bibinfo {author} {\bibfnamefont {J.~T.}\ \bibnamefont {Barreiro}}, \bibinfo {author} {\bibfnamefont {Y.}~\bibnamefont {Wang}}, \ and\ \bibinfo {author} {\bibfnamefont {C.}~\bibnamefont {Wu}},\ }\href {\doibase 10.1103/PhysRevLett.121.167205} {\bibfield  {journal} {\bibinfo  {journal} {Phys. Rev. Lett.}\ }\textbf {\bibinfo {volume} {121}},\ \bibinfo {pages} {167205} (\bibinfo {year} {2018})}\BibitemShut {NoStop}%
\bibitem [{\citenamefont {Read}\ and\ \citenamefont {Sachdev}(1989)}]{Read_1989}%
  \BibitemOpen
  \bibfield  {author} {\bibinfo {author} {\bibfnamefont {N.}~\bibnamefont {Read}}\ and\ \bibinfo {author} {\bibfnamefont {S.}~\bibnamefont {Sachdev}},\ }\href {\doibase https://doi.org/10.1016/0550-3213(89)90061-8} {\bibfield  {journal} {\bibinfo  {journal} {Nuc. Phys.}\ }\textbf {\bibinfo {volume} {B316}},\ \bibinfo {pages} {609} (\bibinfo {year} {1989})}\BibitemShut {NoStop}%
\bibitem [{\citenamefont {Li}\ \emph {et~al.}(1998)\citenamefont {Li}, \citenamefont {Ma}, \citenamefont {Shi},\ and\ \citenamefont {Zhang}}]{Zhang_1998}%
  \BibitemOpen
  \bibfield  {author} {\bibinfo {author} {\bibfnamefont {Y.~Q.}\ \bibnamefont {Li}}, \bibinfo {author} {\bibfnamefont {M.}~\bibnamefont {Ma}}, \bibinfo {author} {\bibfnamefont {D.~N.}\ \bibnamefont {Shi}}, \ and\ \bibinfo {author} {\bibfnamefont {F.~C.}\ \bibnamefont {Zhang}},\ }\href {\doibase 10.1103/PhysRevLett.81.3527} {\bibfield  {journal} {\bibinfo  {journal} {Phys. Rev. Lett.}\ }\textbf {\bibinfo {volume} {81}},\ \bibinfo {pages} {3527} (\bibinfo {year} {1998})}\BibitemShut {NoStop}%
\bibitem [{\citenamefont {Penc}\ \emph {et~al.}(2003)\citenamefont {Penc}, \citenamefont {Mambrini}, \citenamefont {Fazekas},\ and\ \citenamefont {Mila}}]{Mila_2003}%
  \BibitemOpen
  \bibfield  {author} {\bibinfo {author} {\bibfnamefont {K.}~\bibnamefont {Penc}}, \bibinfo {author} {\bibfnamefont {M.}~\bibnamefont {Mambrini}}, \bibinfo {author} {\bibfnamefont {P.}~\bibnamefont {Fazekas}}, \ and\ \bibinfo {author} {\bibfnamefont {F.}~\bibnamefont {Mila}},\ }\href {\doibase 10.1103/PhysRevB.68.012408} {\bibfield  {journal} {\bibinfo  {journal} {Phys. Rev. B}\ }\textbf {\bibinfo {volume} {68}},\ \bibinfo {pages} {012408} (\bibinfo {year} {2003})}\BibitemShut {NoStop}%
\bibitem [{\citenamefont {Greiter}\ and\ \citenamefont {Rachel}(2007)}]{greiter_2007}%
  \BibitemOpen
  \bibfield  {author} {\bibinfo {author} {\bibfnamefont {M.}~\bibnamefont {Greiter}}\ and\ \bibinfo {author} {\bibfnamefont {S.}~\bibnamefont {Rachel}},\ }\href {\doibase 10.1103/PhysRevB.75.184441} {\bibfield  {journal} {\bibinfo  {journal} {Phys. Rev. B}\ }\textbf {\bibinfo {volume} {75}},\ \bibinfo {pages} {184441} (\bibinfo {year} {2007})}\BibitemShut {NoStop}%
\bibitem [{\citenamefont {Corboz}\ \emph {et~al.}(2011)\citenamefont {Corboz}, \citenamefont {L\"auchli}, \citenamefont {Penc}, \citenamefont {Troyer},\ and\ \citenamefont {Mila}}]{Corboz_2011}%
  \BibitemOpen
  \bibfield  {author} {\bibinfo {author} {\bibfnamefont {P.}~\bibnamefont {Corboz}}, \bibinfo {author} {\bibfnamefont {A.~M.}\ \bibnamefont {L\"auchli}}, \bibinfo {author} {\bibfnamefont {K.}~\bibnamefont {Penc}}, \bibinfo {author} {\bibfnamefont {M.}~\bibnamefont {Troyer}}, \ and\ \bibinfo {author} {\bibfnamefont {F.}~\bibnamefont {Mila}},\ }\href {\doibase 10.1103/PhysRevLett.107.215301} {\bibfield  {journal} {\bibinfo  {journal} {Phys. Rev. Lett.}\ }\textbf {\bibinfo {volume} {107}},\ \bibinfo {pages} {215301} (\bibinfo {year} {2011})}\BibitemShut {NoStop}%
\bibitem [{\citenamefont {Bauer}\ \emph {et~al.}(2012)\citenamefont {Bauer}, \citenamefont {Corboz}, \citenamefont {L\"auchli}, \citenamefont {Messio}, \citenamefont {Penc}, \citenamefont {Troyer},\ and\ \citenamefont {Mila}}]{Bauer_2012}%
  \BibitemOpen
  \bibfield  {author} {\bibinfo {author} {\bibfnamefont {B.}~\bibnamefont {Bauer}}, \bibinfo {author} {\bibfnamefont {P.}~\bibnamefont {Corboz}}, \bibinfo {author} {\bibfnamefont {A.~M.}\ \bibnamefont {L\"auchli}}, \bibinfo {author} {\bibfnamefont {L.}~\bibnamefont {Messio}}, \bibinfo {author} {\bibfnamefont {K.}~\bibnamefont {Penc}}, \bibinfo {author} {\bibfnamefont {M.}~\bibnamefont {Troyer}}, \ and\ \bibinfo {author} {\bibfnamefont {F.}~\bibnamefont {Mila}},\ }\href {\doibase 10.1103/PhysRevB.85.125116} {\bibfield  {journal} {\bibinfo  {journal} {Phys. Rev. B}\ }\textbf {\bibinfo {volume} {85}},\ \bibinfo {pages} {125116} (\bibinfo {year} {2012})}\BibitemShut {NoStop}%
\bibitem [{\citenamefont {Bondesan}\ and\ \citenamefont {Quella}(2014)}]{Quella_2014}%
  \BibitemOpen
  \bibfield  {author} {\bibinfo {author} {\bibfnamefont {R.}~\bibnamefont {Bondesan}}\ and\ \bibinfo {author} {\bibfnamefont {T.}~\bibnamefont {Quella}},\ }\href {\doibase https://doi.org/10.1016/j.nuclphysb.2014.07.002} {\bibfield  {journal} {\bibinfo  {journal} {Nuc. Phys.}\ }\textbf {\bibinfo {volume} {B886}},\ \bibinfo {pages} {483} (\bibinfo {year} {2014})}\BibitemShut {NoStop}%
\bibitem [{\citenamefont {Weichselbaum}\ \emph {et~al.}(2018)\citenamefont {Weichselbaum}, \citenamefont {Capponi}, \citenamefont {Lecheminant}, \citenamefont {Tsvelik},\ and\ \citenamefont {L\"auchli}}]{Weichselbaum_2018}%
  \BibitemOpen
  \bibfield  {author} {\bibinfo {author} {\bibfnamefont {A.}~\bibnamefont {Weichselbaum}}, \bibinfo {author} {\bibfnamefont {S.}~\bibnamefont {Capponi}}, \bibinfo {author} {\bibfnamefont {P.}~\bibnamefont {Lecheminant}}, \bibinfo {author} {\bibfnamefont {A.~M.}\ \bibnamefont {Tsvelik}}, \ and\ \bibinfo {author} {\bibfnamefont {A.~M.}\ \bibnamefont {L\"auchli}},\ }\href {\doibase 10.1103/PhysRevB.98.085104} {\bibfield  {journal} {\bibinfo  {journal} {Phys. Rev. B}\ }\textbf {\bibinfo {volume} {98}},\ \bibinfo {pages} {085104} (\bibinfo {year} {2018})}\BibitemShut {NoStop}%
\bibitem [{\citenamefont {Paramekanti}\ and\ \citenamefont {Marston}(2007)}]{Paramekanti_2007}%
  \BibitemOpen
  \bibfield  {author} {\bibinfo {author} {\bibfnamefont {A.}~\bibnamefont {Paramekanti}}\ and\ \bibinfo {author} {\bibfnamefont {J.~B.}\ \bibnamefont {Marston}},\ }\href {\doibase 10.1088/0953-8984/19/12/125215} {\bibfield  {journal} {\bibinfo  {journal} {J. Phys. Condens.}\ }\textbf {\bibinfo {volume} {19}},\ \bibinfo {pages} {125215} (\bibinfo {year} {2007})}\BibitemShut {NoStop}%
\bibitem [{\citenamefont {Corboz}\ \emph {et~al.}(2012{\natexlab{a}})\citenamefont {Corboz}, \citenamefont {Penc}, \citenamefont {Mila},\ and\ \citenamefont {L\"auchli}}]{Corboz_2012}%
  \BibitemOpen
  \bibfield  {author} {\bibinfo {author} {\bibfnamefont {P.}~\bibnamefont {Corboz}}, \bibinfo {author} {\bibfnamefont {K.}~\bibnamefont {Penc}}, \bibinfo {author} {\bibfnamefont {F.}~\bibnamefont {Mila}}, \ and\ \bibinfo {author} {\bibfnamefont {A.~M.}\ \bibnamefont {L\"auchli}},\ }\href {\doibase 10.1103/PhysRevB.86.041106} {\bibfield  {journal} {\bibinfo  {journal} {Phys. Rev. B}\ }\textbf {\bibinfo {volume} {86}},\ \bibinfo {pages} {041106 (R)} (\bibinfo {year} {2012}{\natexlab{a}})}\BibitemShut {NoStop}%
\bibitem [{\citenamefont {Lang}\ \emph {et~al.}(2013)\citenamefont {Lang}, \citenamefont {Meng}, \citenamefont {Muramatsu}, \citenamefont {Wessel},\ and\ \citenamefont {Assaad}}]{Assad_2013}%
  \BibitemOpen
  \bibfield  {author} {\bibinfo {author} {\bibfnamefont {T.~C.}\ \bibnamefont {Lang}}, \bibinfo {author} {\bibfnamefont {Z.~Y.}\ \bibnamefont {Meng}}, \bibinfo {author} {\bibfnamefont {A.}~\bibnamefont {Muramatsu}}, \bibinfo {author} {\bibfnamefont {S.}~\bibnamefont {Wessel}}, \ and\ \bibinfo {author} {\bibfnamefont {F.~F.}\ \bibnamefont {Assaad}},\ }\href {\doibase 10.1103/PhysRevLett.111.066401} {\bibfield  {journal} {\bibinfo  {journal} {Phys. Rev. Lett.}\ }\textbf {\bibinfo {volume} {111}},\ \bibinfo {pages} {066401} (\bibinfo {year} {2013})}\BibitemShut {NoStop}%
\bibitem [{\citenamefont {Nataf}\ \emph {et~al.}(2016{\natexlab{a}})\citenamefont {Nataf}, \citenamefont {Lajk\'o}, \citenamefont {Corboz}, \citenamefont {L\"auchli}, \citenamefont {Penc},\ and\ \citenamefont {Mila}}]{Nataf_honey_2016}%
  \BibitemOpen
  \bibfield  {author} {\bibinfo {author} {\bibfnamefont {P.}~\bibnamefont {Nataf}}, \bibinfo {author} {\bibfnamefont {M.}~\bibnamefont {Lajk\'o}}, \bibinfo {author} {\bibfnamefont {P.}~\bibnamefont {Corboz}}, \bibinfo {author} {\bibfnamefont {A.~M.}\ \bibnamefont {L\"auchli}}, \bibinfo {author} {\bibfnamefont {K.}~\bibnamefont {Penc}}, \ and\ \bibinfo {author} {\bibfnamefont {F.}~\bibnamefont {Mila}},\ }\href {\doibase 10.1103/PhysRevB.93.201113} {\bibfield  {journal} {\bibinfo  {journal} {Phys. Rev. B}\ }\textbf {\bibinfo {volume} {93}},\ \bibinfo {pages} {201113 (R)} (\bibinfo {year} {2016}{\natexlab{a}})}\BibitemShut {NoStop}%
\bibitem [{\citenamefont {Gauth\'e}\ \emph {et~al.}(2020)\citenamefont {Gauth\'e}, \citenamefont {Capponi}, \citenamefont {Mambrini},\ and\ \citenamefont {Poilblanc}}]{Gauthe_2021}%
  \BibitemOpen
  \bibfield  {author} {\bibinfo {author} {\bibfnamefont {O.}~\bibnamefont {Gauth\'e}}, \bibinfo {author} {\bibfnamefont {S.}~\bibnamefont {Capponi}}, \bibinfo {author} {\bibfnamefont {M.}~\bibnamefont {Mambrini}}, \ and\ \bibinfo {author} {\bibfnamefont {D.}~\bibnamefont {Poilblanc}},\ }\href {\doibase 10.1103/PhysRevB.101.205144} {\bibfield  {journal} {\bibinfo  {journal} {Phys. Rev. B}\ }\textbf {\bibinfo {volume} {101}},\ \bibinfo {pages} {205144} (\bibinfo {year} {2020})}\BibitemShut {NoStop}%
\bibitem [{\citenamefont {Anderson}(1952)}]{Anderson_1952}%
  \BibitemOpen
  \bibfield  {author} {\bibinfo {author} {\bibfnamefont {P.~W.}\ \bibnamefont {Anderson}},\ }\href {\doibase 10.1103/PhysRev.86.694} {\bibfield  {journal} {\bibinfo  {journal} {Phys. Rev.}\ }\textbf {\bibinfo {volume} {86}},\ \bibinfo {pages} {694} (\bibinfo {year} {1952})}\BibitemShut {NoStop}%
\bibitem [{\citenamefont {Manousakis}(1991)}]{Manousakis_1991}%
  \BibitemOpen
  \bibfield  {author} {\bibinfo {author} {\bibfnamefont {E.}~\bibnamefont {Manousakis}},\ }\href {\doibase 10.1103/RevModPhys.63.1} {\bibfield  {journal} {\bibinfo  {journal} {Rev. Mod. Phys.}\ }\textbf {\bibinfo {volume} {63}},\ \bibinfo {pages} {1} (\bibinfo {year} {1991})}\BibitemShut {NoStop}%
\bibitem [{\citenamefont {Bernu}\ \emph {et~al.}(1994)\citenamefont {Bernu}, \citenamefont {Lecheminant}, \citenamefont {Lhuillier},\ and\ \citenamefont {Pierre}}]{Lecheminant_1994}%
  \BibitemOpen
  \bibfield  {author} {\bibinfo {author} {\bibfnamefont {B.}~\bibnamefont {Bernu}}, \bibinfo {author} {\bibfnamefont {P.}~\bibnamefont {Lecheminant}}, \bibinfo {author} {\bibfnamefont {C.}~\bibnamefont {Lhuillier}}, \ and\ \bibinfo {author} {\bibfnamefont {L.}~\bibnamefont {Pierre}},\ }\href {\doibase 10.1103/PhysRevB.50.10048} {\bibfield  {journal} {\bibinfo  {journal} {Phys. Rev. B}\ }\textbf {\bibinfo {volume} {50}},\ \bibinfo {pages} {10048} (\bibinfo {year} {1994})}\BibitemShut {NoStop}%
\bibitem [{\citenamefont {Capriotti}\ \emph {et~al.}(1999)\citenamefont {Capriotti}, \citenamefont {Trumper},\ and\ \citenamefont {Sorella}}]{Sorella_1999}%
  \BibitemOpen
  \bibfield  {author} {\bibinfo {author} {\bibfnamefont {L.}~\bibnamefont {Capriotti}}, \bibinfo {author} {\bibfnamefont {A.~E.}\ \bibnamefont {Trumper}}, \ and\ \bibinfo {author} {\bibfnamefont {S.}~\bibnamefont {Sorella}},\ }\href {\doibase 10.1103/PhysRevLett.82.3899} {\bibfield  {journal} {\bibinfo  {journal} {Phys. Rev. Lett.}\ }\textbf {\bibinfo {volume} {82}},\ \bibinfo {pages} {3899} (\bibinfo {year} {1999})}\BibitemShut {NoStop}%
\bibitem [{\citenamefont {Wang}\ and\ \citenamefont {Vishwanath}(2009)}]{Wang_2009}%
  \BibitemOpen
  \bibfield  {author} {\bibinfo {author} {\bibfnamefont {F.}~\bibnamefont {Wang}}\ and\ \bibinfo {author} {\bibfnamefont {A.}~\bibnamefont {Vishwanath}},\ }\href {\doibase 10.1103/PhysRevB.80.064413} {\bibfield  {journal} {\bibinfo  {journal} {Phys. Rev. B}\ }\textbf {\bibinfo {volume} {80}},\ \bibinfo {pages} {064413} (\bibinfo {year} {2009})}\BibitemShut {NoStop}%
\bibitem [{\citenamefont {Hermele}\ and\ \citenamefont {Gurarie}(2011)}]{Hermele_2011}%
  \BibitemOpen
  \bibfield  {author} {\bibinfo {author} {\bibfnamefont {M.}~\bibnamefont {Hermele}}\ and\ \bibinfo {author} {\bibfnamefont {V.}~\bibnamefont {Gurarie}},\ }\href {\doibase 10.1103/PhysRevB.84.174441} {\bibfield  {journal} {\bibinfo  {journal} {Phys. Rev. B}\ }\textbf {\bibinfo {volume} {84}},\ \bibinfo {pages} {174441} (\bibinfo {year} {2011})}\BibitemShut {NoStop}%
\bibitem [{\citenamefont {Corboz}\ \emph {et~al.}(2012{\natexlab{b}})\citenamefont {Corboz}, \citenamefont {Lajk\'o}, \citenamefont {L\"auchli}, \citenamefont {Penc},\ and\ \citenamefont {Mila}}]{Corboz_prx_2012}%
  \BibitemOpen
  \bibfield  {author} {\bibinfo {author} {\bibfnamefont {P.}~\bibnamefont {Corboz}}, \bibinfo {author} {\bibfnamefont {M.}~\bibnamefont {Lajk\'o}}, \bibinfo {author} {\bibfnamefont {A.~M.}\ \bibnamefont {L\"auchli}}, \bibinfo {author} {\bibfnamefont {K.}~\bibnamefont {Penc}}, \ and\ \bibinfo {author} {\bibfnamefont {F.}~\bibnamefont {Mila}},\ }\href {\doibase 10.1103/PhysRevX.2.041013} {\bibfield  {journal} {\bibinfo  {journal} {Phys. Rev. X}\ }\textbf {\bibinfo {volume} {2}},\ \bibinfo {pages} {041013} (\bibinfo {year} {2012}{\natexlab{b}})}\BibitemShut {NoStop}%
\bibitem [{\citenamefont {Lai}(2013)}]{Lai_2013}%
  \BibitemOpen
  \bibfield  {author} {\bibinfo {author} {\bibfnamefont {H.-H.}\ \bibnamefont {Lai}},\ }\href {\doibase 10.1103/PhysRevB.87.205131} {\bibfield  {journal} {\bibinfo  {journal} {Phys. Rev. B}\ }\textbf {\bibinfo {volume} {87}},\ \bibinfo {pages} {205131} (\bibinfo {year} {2013})}\BibitemShut {NoStop}%
\bibitem [{\citenamefont {Chen}\ \emph {et~al.}(2016)\citenamefont {Chen}, \citenamefont {Hazzard}, \citenamefont {Rey},\ and\ \citenamefont {Hermele}}]{Chen_Hazzard_2016}%
  \BibitemOpen
  \bibfield  {author} {\bibinfo {author} {\bibfnamefont {G.}~\bibnamefont {Chen}}, \bibinfo {author} {\bibfnamefont {K.~R.~A.}\ \bibnamefont {Hazzard}}, \bibinfo {author} {\bibfnamefont {A.~M.}\ \bibnamefont {Rey}}, \ and\ \bibinfo {author} {\bibfnamefont {M.}~\bibnamefont {Hermele}},\ }\href {\doibase 10.1103/PhysRevA.93.061601} {\bibfield  {journal} {\bibinfo  {journal} {Phys. Rev. A}\ }\textbf {\bibinfo {volume} {93}},\ \bibinfo {pages} {061601 (R)} (\bibinfo {year} {2016})}\BibitemShut {NoStop}%
\bibitem [{\citenamefont {Nataf}\ \emph {et~al.}(2016{\natexlab{b}})\citenamefont {Nataf}, \citenamefont {Lajk\'o}, \citenamefont {Wietek}, \citenamefont {Penc}, \citenamefont {Mila},\ and\ \citenamefont {L\"auchli}}]{Nataf_chiral_2016}%
  \BibitemOpen
  \bibfield  {author} {\bibinfo {author} {\bibfnamefont {P.}~\bibnamefont {Nataf}}, \bibinfo {author} {\bibfnamefont {M.}~\bibnamefont {Lajk\'o}}, \bibinfo {author} {\bibfnamefont {A.}~\bibnamefont {Wietek}}, \bibinfo {author} {\bibfnamefont {K.}~\bibnamefont {Penc}}, \bibinfo {author} {\bibfnamefont {F.}~\bibnamefont {Mila}}, \ and\ \bibinfo {author} {\bibfnamefont {A.~M.}\ \bibnamefont {L\"auchli}},\ }\href {\doibase 10.1103/PhysRevLett.117.167202} {\bibfield  {journal} {\bibinfo  {journal} {Phys. Rev. Lett.}\ }\textbf {\bibinfo {volume} {117}},\ \bibinfo {pages} {167202} (\bibinfo {year} {2016}{\natexlab{b}})}\BibitemShut {NoStop}%
\bibitem [{\citenamefont {Boos}\ \emph {et~al.}(2020)\citenamefont {Boos}, \citenamefont {Ganahl}, \citenamefont {Lajk\'o}, \citenamefont {Nataf}, \citenamefont {L\"auchli}, \citenamefont {Penc}, \citenamefont {Schmidt},\ and\ \citenamefont {Mila}}]{Boos_2020}%
  \BibitemOpen
  \bibfield  {author} {\bibinfo {author} {\bibfnamefont {C.}~\bibnamefont {Boos}}, \bibinfo {author} {\bibfnamefont {C.~J.}\ \bibnamefont {Ganahl}}, \bibinfo {author} {\bibfnamefont {M.}~\bibnamefont {Lajk\'o}}, \bibinfo {author} {\bibfnamefont {P.}~\bibnamefont {Nataf}}, \bibinfo {author} {\bibfnamefont {A.~M.}\ \bibnamefont {L\"auchli}}, \bibinfo {author} {\bibfnamefont {K.}~\bibnamefont {Penc}}, \bibinfo {author} {\bibfnamefont {K.~P.}\ \bibnamefont {Schmidt}}, \ and\ \bibinfo {author} {\bibfnamefont {F.}~\bibnamefont {Mila}},\ }\href {\doibase 10.1103/PhysRevResearch.2.023098} {\bibfield  {journal} {\bibinfo  {journal} {Phys. Rev. Res.}\ }\textbf {\bibinfo {volume} {2}},\ \bibinfo {pages} {023098} (\bibinfo {year} {2020})}\BibitemShut {NoStop}%
\bibitem [{\citenamefont {Keselman}\ \emph {et~al.}(2020)\citenamefont {Keselman}, \citenamefont {Bauer}, \citenamefont {Xu},\ and\ \citenamefont {Jian}}]{Keselman_2020}%
  \BibitemOpen
  \bibfield  {author} {\bibinfo {author} {\bibfnamefont {A.}~\bibnamefont {Keselman}}, \bibinfo {author} {\bibfnamefont {B.}~\bibnamefont {Bauer}}, \bibinfo {author} {\bibfnamefont {C.}~\bibnamefont {Xu}}, \ and\ \bibinfo {author} {\bibfnamefont {C.-M.}\ \bibnamefont {Jian}},\ }\href {\doibase 10.1103/PhysRevLett.125.117202} {\bibfield  {journal} {\bibinfo  {journal} {Phys. Rev. Lett.}\ }\textbf {\bibinfo {volume} {125}},\ \bibinfo {pages} {117202} (\bibinfo {year} {2020})}\BibitemShut {NoStop}%
\bibitem [{\citenamefont {Chen}\ \emph {et~al.}(2021)\citenamefont {Chen}, \citenamefont {Li}, \citenamefont {Nataf}, \citenamefont {Capponi}, \citenamefont {Mambrini}, \citenamefont {Totsuka}, \citenamefont {Tu}, \citenamefont {Weichselbaum}, \citenamefont {von Delft},\ and\ \citenamefont {Poilblanc}}]{Chen_2021}%
  \BibitemOpen
  \bibfield  {author} {\bibinfo {author} {\bibfnamefont {J.-Y.}\ \bibnamefont {Chen}}, \bibinfo {author} {\bibfnamefont {J.-W.}\ \bibnamefont {Li}}, \bibinfo {author} {\bibfnamefont {P.}~\bibnamefont {Nataf}}, \bibinfo {author} {\bibfnamefont {S.}~\bibnamefont {Capponi}}, \bibinfo {author} {\bibfnamefont {M.}~\bibnamefont {Mambrini}}, \bibinfo {author} {\bibfnamefont {K.}~\bibnamefont {Totsuka}}, \bibinfo {author} {\bibfnamefont {H.-H.}\ \bibnamefont {Tu}}, \bibinfo {author} {\bibfnamefont {A.}~\bibnamefont {Weichselbaum}}, \bibinfo {author} {\bibfnamefont {J.}~\bibnamefont {von Delft}}, \ and\ \bibinfo {author} {\bibfnamefont {D.}~\bibnamefont {Poilblanc}},\ }\href {\doibase 10.1103/PhysRevB.104.235104} {\bibfield  {journal} {\bibinfo  {journal} {Phys. Rev. B}\ }\textbf {\bibinfo {volume} {104}},\ \bibinfo {pages} {235104} (\bibinfo {year} {2021})}\BibitemShut {NoStop}%
\bibitem [{\citenamefont {Jin}\ \emph {et~al.}(2022)\citenamefont {Jin}, \citenamefont {Sun}, \citenamefont {Tu},\ and\ \citenamefont {Zhou}}]{Jin_2022}%
  \BibitemOpen
  \bibfield  {author} {\bibinfo {author} {\bibfnamefont {H.-K.}\ \bibnamefont {Jin}}, \bibinfo {author} {\bibfnamefont {R.-Y.}\ \bibnamefont {Sun}}, \bibinfo {author} {\bibfnamefont {H.-H.}\ \bibnamefont {Tu}}, \ and\ \bibinfo {author} {\bibfnamefont {Y.}~\bibnamefont {Zhou}},\ }\href {\doibase https://doi.org/10.1016/j.scib.2022.03.004} {\bibfield  {journal} {\bibinfo  {journal} {Sci. Bull.}\ }\textbf {\bibinfo {volume} {67}},\ \bibinfo {pages} {918} (\bibinfo {year} {2022})}\BibitemShut {NoStop}%
\bibitem [{\citenamefont {Anderson}(1973)}]{Anderson1973ResonatingVB}%
  \BibitemOpen
  \bibfield  {author} {\bibinfo {author} {\bibfnamefont {P.~W.}\ \bibnamefont {Anderson}},\ }\href@noop {} {\bibfield  {journal} {\bibinfo  {journal} {Mater. Res. Bull.}\ }\textbf {\bibinfo {volume} {8}},\ \bibinfo {pages} {153} (\bibinfo {year} {1973})}\BibitemShut {NoStop}%
\bibitem [{\citenamefont {Sutherland}(1975)}]{sutherland}%
  \BibitemOpen
  \bibfield  {author} {\bibinfo {author} {\bibfnamefont {B.}~\bibnamefont {Sutherland}},\ }\href {\doibase 10.1103/PhysRevB.12.3795} {\bibfield  {journal} {\bibinfo  {journal} {Phys. Rev. B}\ }\textbf {\bibinfo {volume} {12}},\ \bibinfo {pages} {3795} (\bibinfo {year} {1975})}\BibitemShut {NoStop}%
\bibitem [{\citenamefont {Nataf}\ and\ \citenamefont {Mila}(2014)}]{nataf_exact_2014}%
  \BibitemOpen
  \bibfield  {author} {\bibinfo {author} {\bibfnamefont {P.}~\bibnamefont {Nataf}}\ and\ \bibinfo {author} {\bibfnamefont {F.}~\bibnamefont {Mila}},\ }\href {https://link.aps.org/doi/10.1103/PhysRevLett.113.127204} {\bibfield  {journal} {\bibinfo  {journal} {Phys. Rev. Lett.}\ }\textbf {\bibinfo {volume} {113}},\ \bibinfo {pages} {127204} (\bibinfo {year} {2014})}\BibitemShut {NoStop}%
\bibitem [{Note1()}]{Note1}%
  \BibitemOpen
  \bibinfo {note} {Or, more precisely, of its universal enveloping algebra \cite {dixmier1977enveloping}, since it contains some square of generators}\BibitemShut {NoStop}%
\bibitem [{\citenamefont {Wan}\ \emph {et~al.}(2017)\citenamefont {Wan}, \citenamefont {Nataf},\ and\ \citenamefont {Mila}}]{wan_exact_2017}%
  \BibitemOpen
  \bibfield  {author} {\bibinfo {author} {\bibfnamefont {K.}~\bibnamefont {Wan}}, \bibinfo {author} {\bibfnamefont {P.}~\bibnamefont {Nataf}}, \ and\ \bibinfo {author} {\bibfnamefont {F.}~\bibnamefont {Mila}},\ }\href {https://link.aps.org/doi/10.1103/PhysRevB.96.115159} {\bibfield  {journal} {\bibinfo  {journal} {Phys. Rev. B}\ }\textbf {\bibinfo {volume} {96}},\ \bibinfo {pages} {115159} (\bibinfo {year} {2017})}\BibitemShut {NoStop}%
\bibitem [{sup()}]{supp_mat}%
  \BibitemOpen
  \href@noop {} {}\bibinfo {note} {See Supplemental Material at http://link.aps.org/ supplemental/10.1103/PhysRevLett.132.153001 (or below) for some details about representation theory and for some complementary numerical results. The Supplemental Material includes Ref. \cite{Weyl1925Dec,Robinson1961,Hamermesh,chen_grouptheory,young_onsubstitutional_1900,pilch_formulas_1984,cahn_2014}}\BibitemShut {NoStop}%
\bibitem [{Note2()}]{Note2}%
  \BibitemOpen
  \bibinfo {note} {The irreps of $\protect \mathrm {U}(N)$ or $\protect \mathrm {SU}(N)$ are basically the same \cite {supp_mat}; distinction is relevant when we include/exclude representation of generators with non vanishing trace like some combinations of $E_{ii}$.}\BibitemShut {Stop}%
\bibitem [{\citenamefont {Paldus}(2021)}]{Paldus_2021}%
  \BibitemOpen
  \bibfield  {author} {\bibinfo {author} {\bibfnamefont {J.}~\bibnamefont {Paldus}},\ }\href {\doibase 10.1007/s10910-020-01174-7} {\bibfield  {journal} {\bibinfo  {journal} {J. Math. Chem.}\ }\textbf {\bibinfo {volume} {59}},\ \bibinfo {pages} {1} (\bibinfo {year} {2021})}\BibitemShut {NoStop}%
\bibitem [{\citenamefont {Alex}\ \emph {et~al.}(2011)\citenamefont {Alex}, \citenamefont {Kalus}, \citenamefont {Huckleberry},\ and\ \citenamefont {von Delft}}]{alex_2011}%
  \BibitemOpen
  \bibfield  {author} {\bibinfo {author} {\bibfnamefont {A.}~\bibnamefont {Alex}}, \bibinfo {author} {\bibfnamefont {M.}~\bibnamefont {Kalus}}, \bibinfo {author} {\bibfnamefont {A.}~\bibnamefont {Huckleberry}}, \ and\ \bibinfo {author} {\bibfnamefont {J.}~\bibnamefont {von Delft}},\ }\href {\doibase 10.1063/1.3521562} {\bibfield  {journal} {\bibinfo  {journal} {J. Math. Phys. (N.Y.)}\ }\textbf {\bibinfo {volume} {52}},\ \bibinfo {pages} {023507} (\bibinfo {year} {2011})}\BibitemShut {NoStop}%
\bibitem [{\citenamefont {Gelfand}\ and\ \citenamefont {Tsetlin}(1950)}]{Gelfand_1950}%
  \BibitemOpen
  \bibfield  {author} {\bibinfo {author} {\bibfnamefont {I.~M.}\ \bibnamefont {Gelfand}}\ and\ \bibinfo {author} {\bibfnamefont {M.~L.}\ \bibnamefont {Tsetlin}},\ }\href@noop {} {\bibfield  {journal} {\bibinfo  {journal} {Dokl. Akad. Nauk SSSR}\ }\textbf {\bibinfo {volume} {71}},\ \bibinfo {pages} {825} (\bibinfo {year} {1950})}\BibitemShut {NoStop}%
\bibitem [{\citenamefont {Tsunetsugu}\ and\ \citenamefont {Arikawa}(2006)}]{Tsunetsugu_2006}%
  \BibitemOpen
  \bibfield  {author} {\bibinfo {author} {\bibfnamefont {H.}~\bibnamefont {Tsunetsugu}}\ and\ \bibinfo {author} {\bibfnamefont {M.}~\bibnamefont {Arikawa}},\ }\href {\doibase 10.1143/JPSJ.75.083701} {\bibfield  {journal} {\bibinfo  {journal} {J. Phys. Soc. Jpn.}\ }\textbf {\bibinfo {volume} {75}},\ \bibinfo {pages} {083701} (\bibinfo {year} {2006})}\BibitemShut {NoStop}%
\bibitem [{\citenamefont {L\"auchli}\ \emph {et~al.}(2006)\citenamefont {L\"auchli}, \citenamefont {Mila},\ and\ \citenamefont {Penc}}]{Laeuchli_2006}%
  \BibitemOpen
  \bibfield  {author} {\bibinfo {author} {\bibfnamefont {A.}~\bibnamefont {L\"auchli}}, \bibinfo {author} {\bibfnamefont {F.}~\bibnamefont {Mila}}, \ and\ \bibinfo {author} {\bibfnamefont {K.}~\bibnamefont {Penc}},\ }\href {\doibase 10.1103/PhysRevLett.97.087205} {\bibfield  {journal} {\bibinfo  {journal} {Phys. Rev. Lett.}\ }\textbf {\bibinfo {volume} {97}},\ \bibinfo {pages} {087205} (\bibinfo {year} {2006})}\BibitemShut {NoStop}%
\bibitem [{\citenamefont {Szasz}\ \emph {et~al.}(2020)\citenamefont {Szasz}, \citenamefont {Motruk}, \citenamefont {Zaletel},\ and\ \citenamefont {Moore}}]{Szasz_2020}%
  \BibitemOpen
  \bibfield  {author} {\bibinfo {author} {\bibfnamefont {A.}~\bibnamefont {Szasz}}, \bibinfo {author} {\bibfnamefont {J.}~\bibnamefont {Motruk}}, \bibinfo {author} {\bibfnamefont {M.~P.}\ \bibnamefont {Zaletel}}, \ and\ \bibinfo {author} {\bibfnamefont {J.~E.}\ \bibnamefont {Moore}},\ }\href {\doibase 10.1103/PhysRevX.10.021042} {\bibfield  {journal} {\bibinfo  {journal} {Phys. Rev. X}\ }\textbf {\bibinfo {volume} {10}},\ \bibinfo {pages} {021042} (\bibinfo {year} {2020})}\BibitemShut {NoStop}%
\bibitem [{\citenamefont {Feng}\ \emph {et~al.}(2023)\citenamefont {Feng}, \citenamefont {Ibarra-Garc\'{\i}a-Padilla}, \citenamefont {Hazzard}, \citenamefont {Scalettar}, \citenamefont {Zhang},\ and\ \citenamefont {Vitali}}]{Feng_2023}%
  \BibitemOpen
  \bibfield  {author} {\bibinfo {author} {\bibfnamefont {C.}~\bibnamefont {Feng}}, \bibinfo {author} {\bibfnamefont {E.}~\bibnamefont {Ibarra-Garc\'{\i}a-Padilla}}, \bibinfo {author} {\bibfnamefont {K.~R.~A.}\ \bibnamefont {Hazzard}}, \bibinfo {author} {\bibfnamefont {R.}~\bibnamefont {Scalettar}}, \bibinfo {author} {\bibfnamefont {S.}~\bibnamefont {Zhang}}, \ and\ \bibinfo {author} {\bibfnamefont {E.}~\bibnamefont {Vitali}},\ }\href {\doibase 10.1103/PhysRevResearch.5.043267} {\bibfield  {journal} {\bibinfo  {journal} {Phys. Rev. Res.}\ }\textbf {\bibinfo {volume} {5}},\ \bibinfo {pages} {043267} (\bibinfo {year} {2023})}\BibitemShut {NoStop}%
\bibitem [{\citenamefont {Chung}\ and\ \citenamefont {Corboz}(2019)}]{Corboz_2019}%
  \BibitemOpen
  \bibfield  {author} {\bibinfo {author} {\bibfnamefont {S.~S.}\ \bibnamefont {Chung}}\ and\ \bibinfo {author} {\bibfnamefont {P.}~\bibnamefont {Corboz}},\ }\href {\doibase 10.1103/PhysRevB.100.035134} {\bibfield  {journal} {\bibinfo  {journal} {Phys. Rev. B}\ }\textbf {\bibinfo {volume} {100}},\ \bibinfo {pages} {035134} (\bibinfo {year} {2019})}\BibitemShut {NoStop}%
\bibitem [{\citenamefont {Wu}\ and\ \citenamefont {Zhang}(2005)}]{Wu_2005}%
  \BibitemOpen
  \bibfield  {author} {\bibinfo {author} {\bibfnamefont {C.}~\bibnamefont {Wu}}\ and\ \bibinfo {author} {\bibfnamefont {S.-C.}\ \bibnamefont {Zhang}},\ }\href {\doibase 10.1103/PhysRevB.71.155115} {\bibfield  {journal} {\bibinfo  {journal} {Phys. Rev. B}\ }\textbf {\bibinfo {volume} {71}},\ \bibinfo {pages} {155115} (\bibinfo {year} {2005})}\BibitemShut {NoStop}%
\bibitem [{\citenamefont {Flint}\ \emph {et~al.}(2008)\citenamefont {Flint}, \citenamefont {Dzero},\ and\ \citenamefont {Coleman}}]{Coleman_2008}%
  \BibitemOpen
  \bibfield  {author} {\bibinfo {author} {\bibfnamefont {R.}~\bibnamefont {Flint}}, \bibinfo {author} {\bibfnamefont {M.}~\bibnamefont {Dzero}}, \ and\ \bibinfo {author} {\bibfnamefont {P.}~\bibnamefont {Coleman}},\ }\href {\doibase 10.1038/nphys1024} {\bibfield  {journal} {\bibinfo  {journal} {Nat. Phys.}\ }\textbf {\bibinfo {volume} {4}},\ \bibinfo {pages} {643} (\bibinfo {year} {2008})}\BibitemShut {NoStop}%
\bibitem [{\citenamefont {Ramires}(2017)}]{Ramires_2017}%
  \BibitemOpen
  \bibfield  {author} {\bibinfo {author} {\bibfnamefont {A.}~\bibnamefont {Ramires}},\ }\href {\doibase 10.1103/PhysRevA.96.043604} {\bibfield  {journal} {\bibinfo  {journal} {Phys. Rev. A}\ }\textbf {\bibinfo {volume} {96}},\ \bibinfo {pages} {043604} (\bibinfo {year} {2017})}\BibitemShut {NoStop}%
\bibitem [{\citenamefont {Molev}(1999)}]{Molev_1999}%
  \BibitemOpen
  \bibfield  {author} {\bibinfo {author} {\bibfnamefont {A.~I.}\ \bibnamefont {Molev}},\ }\href {\doibase 10.1007/s002200050570} {\bibfield  {journal} {\bibinfo  {journal} {Commun. Math. Phys.}\ }\textbf {\bibinfo {volume} {201}},\ \bibinfo {pages} {591} (\bibinfo {year} {1999})}\BibitemShut {NoStop}%
\bibitem [{\citenamefont {{Molev}}(2002)}]{Molev_2002}%
  \BibitemOpen
  \bibfield  {author} {\bibinfo {author} {\bibfnamefont {A.~I.}\ \bibnamefont {{Molev}}},\ }\href {\doibase 10.48550/arXiv.math/0211289} {\bibfield  {journal} {\bibinfo  {journal} {arXiv}\ ,\ \bibinfo {eid} {math/0211289}} (\bibinfo {year} {2002})},\ \Eprint {http://arxiv.org/abs/math/0211289} {arXiv:math/0211289} \BibitemShut {NoStop}%
\bibitem [{\citenamefont {Nataf}\ and\ \citenamefont {Mila}(2018)}]{nataf_density_2018}%
  \BibitemOpen
  \bibfield  {author} {\bibinfo {author} {\bibfnamefont {P.}~\bibnamefont {Nataf}}\ and\ \bibinfo {author} {\bibfnamefont {F.}~\bibnamefont {Mila}},\ }\href {https://link.aps.org/doi/10.1103/PhysRevB.97.134420} {\bibfield  {journal} {\bibinfo  {journal} {Phys. Rev. B}\ }\textbf {\bibinfo {volume} {97}},\ \bibinfo {pages} {134420} (\bibinfo {year} {2018})}\BibitemShut {NoStop}%
\bibitem [{\citenamefont {Gozel}\ \emph {et~al.}(2020)\citenamefont {Gozel}, \citenamefont {Nataf},\ and\ \citenamefont {Mila}}]{gozel_2020}%
  \BibitemOpen
  \bibfield  {author} {\bibinfo {author} {\bibfnamefont {S.}~\bibnamefont {Gozel}}, \bibinfo {author} {\bibfnamefont {P.}~\bibnamefont {Nataf}}, \ and\ \bibinfo {author} {\bibfnamefont {F.}~\bibnamefont {Mila}},\ }\href {\doibase 10.1103/PhysRevLett.125.057202} {\bibfield  {journal} {\bibinfo  {journal} {Phys. Rev. Lett.}\ }\textbf {\bibinfo {volume} {125}},\ \bibinfo {pages} {057202} (\bibinfo {year} {2020})}\BibitemShut {NoStop}%
\bibitem [{\citenamefont {Vilenkin}\ and\ \citenamefont {Klimyk}(1992)}]{Vilenkin_vol3}%
  \BibitemOpen
  \bibfield  {author} {\bibinfo {author} {\bibfnamefont {N.~J.}\ \bibnamefont {Vilenkin}}\ and\ \bibinfo {author} {\bibfnamefont {A.~U.}\ \bibnamefont {Klimyk}},\ }\href@noop {} {\emph {\bibinfo {title} {Representation of Lie Groups and Special Functions, Vol. 3.}}}\ (\bibinfo  {publisher} {Kluwer Academic Publishers},\ \bibinfo {year} {1992,})\BibitemShut {NoStop}%
\bibitem [{\citenamefont {Dixmier}(1977)}]{dixmier1977enveloping}%
  \BibitemOpen
  \bibfield  {author} {\bibinfo {author} {\bibfnamefont {J.}~\bibnamefont {Dixmier}},\ }\href {https://books.google.fr/books?id=Dl4ljyqGG9wC} {\emph {\bibinfo {title} {Enveloping Algebras}}},\ North-Holland Mathematical library\ (\bibinfo  {publisher} {North-Holland Publishing Company, Amsterdam},\ \bibinfo {year} {1977})\BibitemShut {NoStop}%
\bibitem [{\citenamefont {Weyl}(1925)}]{Weyl1925Dec}%
  \BibitemOpen
  \bibfield  {author} {\bibinfo {author} {\bibfnamefont {H.}~\bibnamefont {Weyl}},\ }\href {\doibase 10.1007/BF01506234} {\bibfield  {journal} {\bibinfo  {journal} {Math. Z.}\ }\textbf {\bibinfo {volume} {23}},\ \bibinfo {pages} {271} (\bibinfo {year} {1925})}\BibitemShut {NoStop}%
\bibitem [{\citenamefont {de~B.~Robinson}(1961)}]{Robinson1961}%
  \BibitemOpen
  \bibfield  {author} {\bibinfo {author} {\bibfnamefont {G.}~\bibnamefont {de~B.~Robinson}},\ }\href@noop {} {\emph {\bibinfo {title} {{Representation Theory of the Symmetric Group}}}}\ (\bibinfo  {publisher} {Edinburgh: University Press, Edinburgh},\ \bibinfo {year} {1961})\BibitemShut {NoStop}%
\bibitem [{\citenamefont {Hamermesh}(1989)}]{Hamermesh}%
  \BibitemOpen
  \bibfield  {author} {\bibinfo {author} {\bibfnamefont {M.}~\bibnamefont {Hamermesh}},\ }\href {https://cds.cern.ch/record/1123140} {\emph {\bibinfo {title} {Group Theory and Its Application to Physical Problems, Corr. ed.}}}\ (\bibinfo  {publisher} {Dover},\ \bibinfo {address} {New York, NY},\ \bibinfo {year} {1989})\BibitemShut {NoStop}%
\bibitem [{\citenamefont {Chen}\ \emph {et~al.}(2002)\citenamefont {Chen}, \citenamefont {Ping},\ and\ \citenamefont {Wang}}]{chen_grouptheory}%
  \BibitemOpen
  \bibfield  {author} {\bibinfo {author} {\bibfnamefont {J.-Q.}\ \bibnamefont {Chen}}, \bibinfo {author} {\bibfnamefont {J.}~\bibnamefont {Ping}}, \ and\ \bibinfo {author} {\bibfnamefont {F.}~\bibnamefont {Wang}},\ }\href@noop {} {\emph {\bibinfo {title} {Group Representation Theory for Physicists}}}\ (\bibinfo  {publisher} {World Scientific Publishing Company, Singapore},\ \bibinfo {year} {2002})\BibitemShut {NoStop}%
\bibitem [{\citenamefont {Young}(1900)}]{young_onsubstitutional_1900}%
  \BibitemOpen
  \bibfield  {author} {\bibinfo {author} {\bibfnamefont {A.}~\bibnamefont {Young}},\ }\href {https://londmathsoc.onlinelibrary.wiley.com/doi/abs/10.1112/plms/s1-33.1.97} {\bibfield  {journal} {\bibinfo  {journal} {Proc. London Math. Soc.}\ }\textbf {\bibinfo {volume} {s1-33}},\ \bibinfo {pages} {97} (\bibinfo {year} {1900})}\BibitemShut {NoStop}%
\bibitem [{\citenamefont {Pilch}\ and\ \citenamefont {Schellekens}(1984)}]{pilch_formulas_1984}%
  \BibitemOpen
  \bibfield  {author} {\bibinfo {author} {\bibfnamefont {K.}~\bibnamefont {Pilch}}\ and\ \bibinfo {author} {\bibfnamefont {A.~N.}\ \bibnamefont {Schellekens}},\ }\href {https://doi.org/10.1063/1.526101} {\bibfield  {journal} {\bibinfo  {journal} {J. Math. Phys. (N.Y.)}\ }\textbf {\bibinfo {volume} {25}},\ \bibinfo {pages} {3455} (\bibinfo {year} {1984})}\BibitemShut {NoStop}%
\bibitem [{\citenamefont {Cahn}(2014)}]{cahn_2014}%
  \BibitemOpen
  \bibfield  {author} {\bibinfo {author} {\bibfnamefont {R.}~\bibnamefont {Cahn}},\ }\href {https://books.google.fr/books?id=393zAwAAQBAJ} {\emph {\bibinfo {title} {Semi-Simple Lie Algebras and Their Representations}}}\ (\bibinfo  {publisher} {Dover Publications, New York,},\ \bibinfo {year} {2014})\BibitemShut {NoStop}%
\end{thebibliography}%

\clearpage

\onecolumngrid

\begin{center}
	\textbf{\large Exact Diagonalization of $\mathrm{SU}(N)$ Fermi-Hubbard Models\\ [.3cm] -- Supplemental Material --}\\[.4cm]
	Thomas Botzung$^{1}$, and  Pierre Nataf$^{1}$ \\[.1cm]
	{\itshape ${}^1$ Laboratoire de Physique et Mod\'elisation des Milieux Condens\'es, Universit\'e Grenoble Alpes and CNRS, 25 avenue des Martyrs, 38042 Grenoble, France}
	(Dated: \today)\\[1cm]
\end{center}

\setcounter{section}{0}
\setcounter{equation}{0}
\setcounter{figure}{0}
\setcounter{table}{0}
\setcounter{page}{1}

\renewcommand{\theequation}{S\arabic{equation}}
\renewcommand{\thefigure}{S\arabic{figure}}
\renewcommand{\bibnumfmt}[1]{[S#1]}

 In this supplemental material, we first provide representation theory background to detail/support some claims and formula written in the main article and then we show complementary numerical results obtained via ED. 
 
In Sec.~\ref{basis}, we review the basis of the irreps of U($N$) and SU($N$) and in Sec.~\ref{generators} the construction of the finite-dimensional representations of the generators of  U($N$) and SU($N$).
Then, in Sec.~\ref{factorization}, we detail the color factorization at work in our ED procedure, before focusing on the construction of the SU($N$) singlets basis in \,Sec.~\ref{singlets}.
Next, in Sec.~\ref{largeU}, we show how to recover the Young rules used in the diagonalization of the SU($N$) Heisenberg Model (HM) from the Gelfand-Tsetlin rules in the large U limit of the SU($N$) Fermi-Hubbard Model (FHM).
In the final subsection of the group theory part, i.e in Sec.~\ref{symplectic}, we show the dimensions of the symplectic  Sp($N$) singlets subspace for some specific parameters.

Then, we move to the complementary numerical part. In  Sec.~\ref{sec: su5}, we focus on SU($5$) on the square lattice with L=10 sites. For this specific case, we show some energy spectra and we detail  the fitting procedure of the charge gap that we have used for all the systems under consideration in our manuscript.  Then, in Sec.~\ref{sec: su2}, we consider the case SU($2$) on a triangular lattice with $L=12$ and $L=16$ sites which illustrate the finite size effects on the spin/charge gaps. Finally, in Sec.~\ref{sec: correlation}, we show the evolution of real space correlations while increasing the interaction $U$ for SU($3$) on the L=12 triangular lattice and  for SU($5$) on the square lattice.
\section{Representations of the (special) unitary groups}
\label{sec: suN}
\subsection{Basis of the irreps of SU($N$)}
\label{basis}
\begin{figure}[t]
    \centering
    \includegraphics[width=1\textwidth]{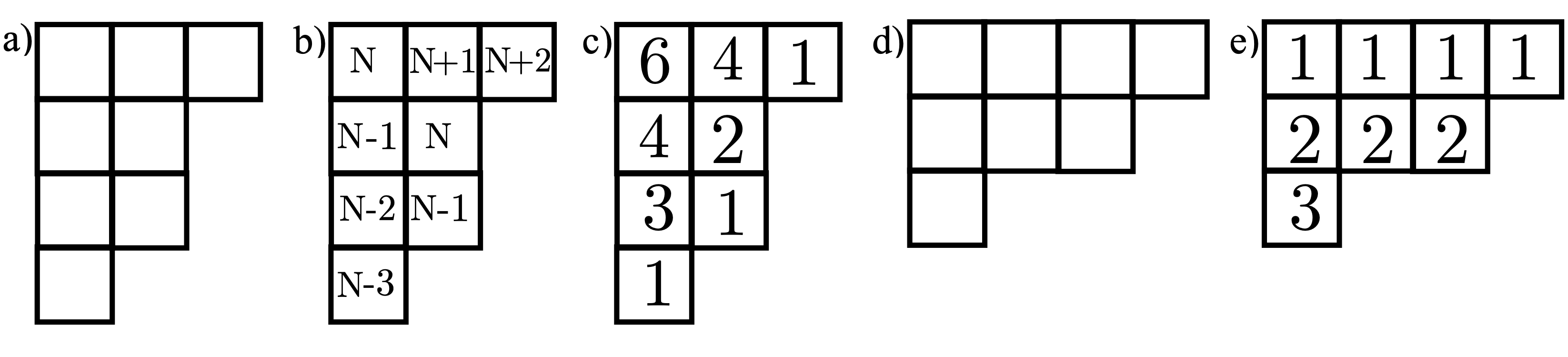}
    \caption{(a) Typical Young Tableau or Young Diagram (here $\alpha=[3,2,2,1])$; (b) quantities $d_{i,N}$ appearing in the calculation of the dimension $d^{\alpha}_N$ of an U($N$) or SU($N$) irrep (cf Eq. \eqref{supp_equation_dim});
    (c)Hook lengths $l_i$ which also appear in the calculation $d^{\alpha}_N$, d) $\bar{\alpha}=[4,3,1]$: transpose of $\alpha=[3,2,2,1] $, e) Highest Weight State associated with $\bar{\alpha}=[4,3,1]$. See text for details.}
    \label{young_diagram}
\end{figure}

Each irreducible representation ({\it irrep}) of the special unitary group $SU(N)$ (or of the unitary group U($N$)) is labeled by a Young Diagram (or {\it shape})  $\alpha=[\alpha_1,\alpha_2,...,\alpha_k]$ , with $k$ the number of rows of the diagram ($1\leq k \leq N$) and where the lengths of the rows $\alpha_j$ satisfy $\alpha_1\geq\alpha_2\geq...\geq \alpha_k \geq 1$ (see Fig.\ref{young_diagram} a) for an example).

\begin{figure}[t]
    \centering
    \includegraphics[width=1\textwidth]{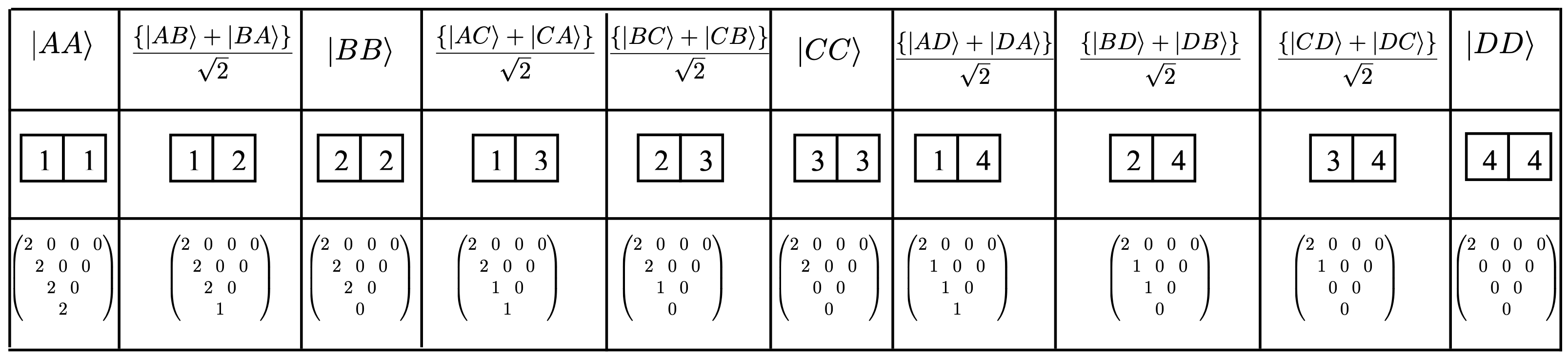}
    \caption{Ten-dimensional basis of the $U(4)$ or $SU(4)$ fully symmetric irrep $\alpha=[2]$. Top row: wave-functions written with $N=4$ colors, $A,B,C$ and $D$. Middle row: same states written with semi-standard Young Tableaux (ssYT), after the mapping $A \rightarrow 1$, $B \rightarrow 2$, $C \rightarrow 3$ and $D \rightarrow 4$; Bottom row: equivalent Gelfand Tsetlin patterns. Cf text for details.}
    \label{sketch_basis}
\end{figure}

The dimension $d^{\alpha}_N$ of a  $SU(N)$ irrep of shape $\alpha$ can be calculated \cite{Weyl1925Dec,Robinson1961}
 simply from the shape $\alpha$ as $d^{\alpha}_N=\prod_{i=1}^{M} (d_{i,N}/ l_i )$, with $M$ the number of boxes, with $d_{i,N}=N+\gamma_i$, where  $\gamma_i$ is the algebraic distance from the $i^{th}$ box to the main diagonal, counted positively (resp. negatively) for a box above (below) the diagonal  (see Fig.\ref{young_diagram} b). The hook lengths $l_i$ of a box are defined as the number of boxes on the same row at the right plus the number of boxes in the same column below plus the box itself (see Fig.\ref{young_diagram} c).
An equivalent and useful (especially when the number of boxes increases) formula for $d^{\alpha}_N$ is \cite{alex_2011}:
\begin{equation}
\label{supp_equation_dim}
d^{\alpha}_N = \prod_{1 \leq j < j' \leq N} \Big{(} 1+\frac{\alpha_j-\alpha_{j'}}{j'-j} \Big{)},
\end{equation}
where $\alpha_j=0$ for $k < j \leq N$ (when $k<N$, we {\it add} rows of length $0$ to have $N$ entries for the Young Diagram $\alpha=[\alpha_1,\alpha_2,...,\alpha_k] \rightarrow [\alpha_1,\alpha_2,...,\alpha_k,0, \cdots, 0]$).
For instance, for $N=4$ colors, for the fully symmetric irrep $\lambda=[2]$ represented as 
$
\begin{Young}
& \cr
\end{Young}
$
one has $d^{\alpha}_N=10$. 
 If we call $A,B,C$ and $D$ the four colors, 
the $10$ orthonormal (and symmetric) basis states of the irrep $[2]$ can be chosen as:
\begin{align}
\label{symmetric_states}
\{ |AA\rangle, |BB\rangle&,|CC\rangle,|DD\rangle\nonumber\\
\frac{1}{\sqrt{2}}\{|AB\rangle+|BA\rangle\}&,\frac{1}{\sqrt{2}}\{|AC\rangle+|CA\rangle\} \nonumber\\
\frac{1}{\sqrt{2}}\{|AD\rangle+|DA\rangle\}&,\frac{1}{\sqrt{2}}\{|BC\rangle+|CB\rangle\} \nonumber\\
\frac{1}{\sqrt{2}}\{|BD\rangle+|DB\rangle\}&,\frac{1}{\sqrt{2}}\{|CD\rangle+|DC\rangle\} \}
\end{align}
More generally, an orthonormal basis of an irrep $\alpha$ can be labelled by the set of semi-standard Young tableaux \cite{Hamermesh,chen_grouptheory}:
a ssYT is filled up with integers numbers from 1 to $N$ in non-descending order from left to right in any row (repetitions allowed), and in strictly ascending order (repetitions not allowed) from top to bottom in any column. In Fig. \ref{sketch_basis} we give the correspondence between the basis written in Eq. \eqref{symmetric_states} and the set of ssYTs of shape $\alpha=[2]=
\begin{Young}
& \cr
\end{Young}
$ , after the mapping $A \rightarrow 1$, $B \rightarrow 2$, $C \rightarrow 3$ and $D \rightarrow 4$.

We also provide the correspondence between the set of ssYTs and the set of Gelfand-Tsetlin patterns (GT-patterns). GT-patterns are triangular arrangements of integers, denoted by $M = (m_{k,l})$, fully equivalent to the ssYTs, with the structure \cite{Gelfand_1950,Vilenkin_vol3,alex_2011}:

\begin{equation}
M=
\begin{pmatrix}
m_{1,N}\hspace{-0.15cm}&\hspace{-0.15cm}&\hspace{-0.5cm}m_{2,N}\hspace{-0.15cm}&\hspace{-0.15cm}\cdots&\hspace{-0.15cm}\cdots &\hspace{-0.15cm}m_{N-1,N}\hspace{-0.15cm}&\hspace{-0.15cm}&\hspace{-0.5cm}m_{N,N}\\
\hspace{-0.15cm}&\hspace{-0.15cm}m_{1,N-1}\hspace{-0.15cm}&\hspace{-0.15cm}&\hspace{-0.15cm}\cdots\hspace{-0.15cm}&\hspace{-0.15cm}\cdots \hspace{-0.15cm}&\hspace{-0.15cm}\cdots &\hspace{-0.15cm}m_{N-1,N-1}\hspace{-0.15cm}& \\
\hspace{-0.15cm}&\cdots\hspace{-0.15cm}&\hspace{-0.15cm}&\hspace{-0.15cm}\cdots\hspace{-0.15cm}&\hspace{-0.15cm}\cdots \hspace{-0.5cm}&\hspace{-0.5cm}\cdots&\hspace{-0.75cm}\cdots\hspace{-0.15cm}\cdots \hspace{-0.15cm}& \\
\hspace{-0.15cm}&\hspace{0.5cm}\cdots\hspace{-0.15cm}&\hspace{-0.15cm}&\hspace{-0.15cm}\cdots\hspace{-0.15cm}&\hspace{-0.15cm}\cdots \hspace{-0.5cm}&\hspace{-0.5cm}\cdots&\hspace{-2cm}\cdots\hspace{-0.15cm}\hspace{-0.15cm}& \\
\hspace{-0.15cm}&\hspace{-0.15cm}&&\hspace{-0.15cm}m_{1,2}\hspace{-0.15cm}&\hspace{-0.15cm}&\hspace{-0.75cm}m_{2,2}\hspace{-0.15cm}&&\hspace{-0.15cm}&\\
&\hspace{-0.15cm}&\hspace{-0.15cm}&&m_{1,1}\hspace{-0.15cm}&\hspace{-0.15cm}&\hspace{-0.15cm}&
\end{pmatrix},
\end{equation}
where the lengths of the rows of the irrep $\alpha$ appears in the first line:  $m_{j,N}=\alpha_j$ for  $1\leq j \leq N$, and where the non negative integer coefficients
$m_{k,l}$ are constrained by the betweenness condition:
\begin{equation}
m_{j,p} \geq m_{j,p-1} \geq m_{j+1,p}, \hspace {2 cm} \text{for} \,\,  1 \leq j < p \leq N.
\end{equation}
The systematic way to convert  a GT-pattern M to a ssYT is well explained in the Appendix A of \cite{alex_2011}.

\subsection{Generators of SU($N$) and their representations}
\label{generators}
We first describe the $N^2$ generators of the unitary group U($N$).
They satisfy the following commutation relations ($1 \leq i,j,k,l \leq N$):

\begin{equation}
\label{supp_commutation}
[E_{ij},E_{kl}]=\delta_{jk}E_{il}-\delta_{li}E_{kj}.
\end{equation}
For a given irrep $\alpha$, the operators $E_{p-1 p}$, $E_{p p-1}$ ($2\leq p \leq N$) and $E_{p p}$  ($1\leq p \leq N$) behave on a  GT-pattern $M\equiv \vert M \rangle$ (seen as a basis state in $\alpha$) as  \cite{Gelfand_1950,Vilenkin_vol3}:
\begin{align}
E_{p-1 p} \vert M \rangle &=\sum_{j=1}^{p-1} a^j_{p-1}  \vert M^{+j}_{p-1}\rangle \label{Epmoins1p_rule} \\
E_{p p-1} \vert M \rangle &=\sum_{j=1}^{p-1} b^j_{p-1}  \vert M^{-j}_{p-1}\rangle \label{Eppmoins1_rule}\\
E_{p p} \vert M \rangle &=\Big{(}\sum_{i=1}^{p}m_{i,p}-\sum_{j=1}^{p-1} m_{j,p-1} \Big{)}\vert M\rangle, 
\end{align}
where the basis states $\vert M^{\pm j}_{p-1}\rangle $ designate GT-patterns equal to $M$ except for the coefficient $m_{j,p-1}$ replaced by  $m_{j,p-1}\pm1$, and where the coefficients $a^j_{p-1}$ and $b^j_{p-1}$
read:
\begin{align}
\label{mat_coeff}
a^j_{p-1}&=  \left | \frac{\prod_{i=1}^p (l_{i,p}-l_{j,p-1})\prod_{i=1}^{p-2} (l_{i,p-2}-l_{j,p-1}-1)}{ \prod_{i\neq j} (l_{i,p-1}-l_{j,p-1})\prod_{i\neq j} (l_{i,p-1}-l_{j,p-1}-1)} \right | ^{1/2},\\
b^j_{p-1}&=  \left | \frac{\prod_{i=1}^p (l_{i,p}-l_{j,p-1}+1)\prod_{i=1}^{p-2} (l_{i,p-2}-l_{j,p-1})}{ \prod_{i\neq j} (l_{i,p-1}-l_{j,p-1})\prod_{i\neq j} (l_{i,p-1}-l_{j,p-1}+1)} \right | ^{1/2}, \label{mat_coeff2}
\end{align}
with $l_{p,q}=m_{p,q}-p$.\\
In the main text, we give a fully equivalent formulation in terms of ssYT. 
Moreover, to get the matrices representing the other generators, we use iteratively the commutation relations Eqs. \eqref{supp_commutation}: the generators $E_{p p+j} \, (j>1)$ are deduced from the generators $E_{p p+q}$ ($q<j$) through $E_{p p+j}=[E_{p p+1},E_{p+1 p+j}]$. Secondly, the matrices representing the operators $E_{ij}$ are hermitian conjugate of the ones representing the operators $E_{ji}$: $E_{ij}=E_{ji}^{\dag}$ (with some abuse of notation: the generators should not be confused with the matrices representing them).

As for the $N^2-1$ generators of the unitary group SU($N$) which is included in U($N$), they can be written as linear superpositions of the
generators $E_{ij}$ ($1 \leq i,j,\leq N$). The only additional requirement is that the $N-1$ {\it commuting ones}, i.e spanning the Cartan subalgebra \cite{cahn_2014},
should be selected to be traceless, in order that the exponential of their linear superpositions has determinant equal to one.
One possible choice is \cite{chen_grouptheory}: 
\begin{equation}
H_i=E_{ii}-(1/N) (\sum_j E_{jj}) \hspace{1.5cm} \text{for} \hspace{.5cm} i=1\cdots N,
\end{equation}
 among which only $N-1$ are independents since $\sum_i H_i=0$.

 Another choice for the generators of SU($N$)  is given by:  
 
 \begin{equation}
 \mathcal{Q}_r=(\lambda^r)_{\sigma' \sigma} E_{\sigma'  \sigma},  \hspace{1.5cm}\text{where} \hspace{.5cm} 1 \leq \sigma,\sigma' \leq N,
 \end{equation}
  and where the $N^2-1$ matrices $\lambda^r$ are the hermitian and traceless $N \times N$ generalized Gell-Mann  matrices.
 Each of the $N^2-1$ indices $r$ designates a set of two entries: $r \equiv \{b, (j,k) \}$,  where $b=s,a,d$ means either symmetric, antisymmetric or diagonal and where $(j,k)$ is a couple of positive unequal (when $b=s,a$) or equal (when $b=d$) indices not larger than N.
 When, $b=s,a$, one has: $\lambda^{\{s,(j,k)\}}=e_{j,k}+e_{k,j}$ and $\lambda^{\{a,(j,k)\}}=i(-e_{j,k}+e_{k,j})$,
 where $e_{p,l}$ is a $N \times N$ matrix full of zeros except at position $(p,l)$ where it is equal to 1.
 When $b=d$, $j=k<N$ and $\lambda^{\{d,(j,j)\}}=\sqrt{\frac{2}{j(j+1)}} (\sum_{q=1}^j e_{q,q}-j e_{j+1,j+1})$.
 From the commutation relations in Eq. \eqref{supp_commutation}, and the matrix coefficients of $\lambda^r$, one can also get the structure constants $f_{\alpha \beta \gamma}$
which are such that  $[\mathcal{Q}_{r},\mathcal{Q}_{s}]=i f_{rst} \mathcal{Q}_{t}$.

\subsection{Color factorization for $M$ SU($N$) fermions on a $L$-sites lattice}
\label{factorization}
We explain and illustrate here the tensorial decomposition of the Hilbert space $\mathcal{H}^{M,N}_L$, which describes $M$ SU($N$) fermions on L sites interacting through the SU($N$) Fermi-Hubbard Models (FHM) given in Eq. (1) of the main text and written in Eq. \eqref{supp_Hamiltonian} below. \\
For $L=1$ site, the M-fermions wave-functions are fully antisymmetric in the exchange of colors, and belong to the M-boxes single column irrep $\alpha=[1,1,..\alpha_M=1]$, and its dimension  $D^{M,N}_{L=1}=\binom{N}{M}$ simply corresponds to all the ways to choose $M$ colors among N. The wave-functions are simply: $\Pi_{j=1}^M c^{\dag}_{\sigma_j} \vert 0 \rangle $, where the $\sigma_j$ ($j=1 \cdots M$) stand for the $M$ different colors.

 More generally, for $L$ sites and $M$  $\mathrm{SU}(N)$ fermions, the Hilbert space $\mathcal{H}^{M,N}_L$ has dimension $D^{M,N}_L$, which is:
  \begin{equation}
\label{supp_decomposition}
D^{M,N}_L=\sum_{\boldmath{\alpha}} h^{\bar{\alpha}}_L  \prod_{i=1}^L \binom{N}{\bar{\alpha}_i}=\sum_{\boldmath{\alpha}} d^{\alpha}_N  d^{\bar{\alpha}}_L=\sum_{\boldmath{\alpha}} h^{\alpha}_N  \prod_{i=1}^N \binom{L}{\alpha_i},
\end{equation}
where the sums run over all the $M-$boxes YT $\alpha$ of maximum $L$ columns and $N$ rows.
Note that we have added another part to the Eq. (3) of the main text, since there is a duality between the site and the color indices.

$\bar{\alpha}=[\bar{\alpha}_1,...,\bar{\alpha}_L]$ is defined as the {\it transpose}  YT of $\alpha$, transforming rows into columns (see Fig.\ref{young_diagram} a) for an example).
In the LHS of Eq.~\eqref{decomposition}, $\bar{\alpha}$ is a partition of $M$ into $L$ integers not larger than $N$. It can be seen as a
 distribution of fermions: $\bar{\alpha}_j$ being the number of fermions (necessarily $\leq N$) on site $j$ for $1 \leq j \leq L$,
$ \prod_{j=1}^L \binom{N}{\bar{\alpha}_j}$ is the number of states for such a distribution, because for each site $j$ with $\bar{\alpha}_j$ fermions on it, one has  $\binom{N}{\bar{\alpha}_j}$ different possible states (cf case $L=1$ seen above).
The factor  $h^{\bar{\alpha}}_L$, defined as $h^{\bar{\alpha}}_L=L!/ \prod_{k=0}^N (n_k^{\bar{\alpha}})!$, where $n_k^{\bar{\alpha}}=\text{Cardinal}\{j \in \llbracket 1;L \rrbracket / \bar{\alpha}_j=k\}$, is the number of distributions corresponding to
a given partition $\bar{\alpha}$ while permuting the $\bar{\alpha}_j$ (or the site indices $j$) for $1 \leq j \leq L$.
For instance, for $M=4$ particles, $N\geq2$ and $L=6$ sites for the irrep $\alpha=[3,1]$ , the associate transpose shape is $\bar{\alpha}=[2,1,1]\equiv [2,1,1,0,0,0]$ (we add zeros so that $\bar{\alpha}$ has $L$ entries) and corresponds to a distribution of particles $\{2,1,1,0,0,0\}$, and there are $6!/(1!2! 3!)=60$ ways
to select 3 sites without fermion, 2 sites with one fermion and 1 site with two fermions. For each such combination, one has $\prod_{i=1}^L \binom{N}{\bar{\alpha}_i}=\binom{N}{1}^2 \times \binom{N}{2} $ possible states.

In the middle term of Eq.~\eqref{decomposition}, $d^{\alpha}_N $ (resp. $d^{\bar{\alpha}}_L$) stands for the dimension of the $\mathrm{SU}(N)$ irrep $\alpha$ (resp. the $\mathrm{U}(L)$ irrep $\bar{\alpha}$), cf section \ref{basis}.

Finally, in the RHS of Eq.~\eqref{decomposition}, one can interpret each partition $\alpha$ as a distribution of colors and there are  $h^{\alpha}_N=N!/ \prod_{k=0}^N (n_k^{\alpha})!$ ways to have such a distribution of colors; for instance  for $M=4$ particles, $N=4$ colors  and $L=6$ sites, the irrep $\alpha=[3,1] \equiv [3,1,0,0]$ corresponds to the partition $\{3,1,0,0\}$ (we add zeros so that $\alpha$ has $N$ entries). Then, there are $h^{[3,1,0,0]}_4=12$ ways to select 3 identical colors, and one color among $N=4$ (for instance three colors $A$, one $B$ and zero $C$ or $D$).
For each such a distribution, one has $\prod_{i=1}^N \binom{L}{\alpha_i}$ ways to choose the sites to put the colored fermions on. For instance, for $M=N=4$, $L=6$ and  $\alpha=[3,1,0,0]$, one has $\prod_{i=1}^4 \binom{6}{\alpha_i}=120$ ways 
to select three sites among $6$ to put the fermions with color $A$, one site among $6$ to put the color $B$, etc...
In the array  \ref{table_dec}, we give the quantities appearing in  Eq. \eqref{decomposition} for $N=M=4$ and $L=6$ for all the shapes $\alpha$.
\begin{table*}[ht]
        \centering
        \begin{tabular}{|c|c|c|c|c|c|}
         \hline
        $\alpha$ &[1,1,1,1]&[2,1,1,0]&[2,2,0,0]&[3,1,0,0]&[4,0,0,0] \\
        \hline
         $\bar{\alpha}$ &[4,0,0,0,0,0]&[3,1,0,0,0,0]&[2,2,0,0,0,0]&[2,1,1,0,0,0]&[1,1,1,1,0,0] \\
          \hline
            $d^{\alpha}_N$ &1&15&20&45&35 \\
            \hline
            &&&&&\\
                 
  $d^{\bar{\alpha}}_L$&126&210&105&105&15\\
 \hline
            &&&&&\\
  $h^{\bar{\alpha}}_L$&6&30&15&60&15 \\
  \hline
            &&&&&\\
                  $\prod_{i=1}^L \binom{N}{\bar{\alpha}_i}$&1&16&36&96&256 \\
\hline
   &&&&&\\
          $h^{\alpha}_N$&1&12&6&12&4\\
                \hline
                 &&&&&\\
               $\prod_{i=1}^N \binom{L}{\alpha_i}$&$d^{U(1)}_{L,M,N}=1296$&540&225&120&15 \\
\hline
        \end{tabular}
        \caption{Quantities appearing in the calculation of the dimension of the Hilbert space  $\mathcal{H}^{M,N}_L$, shown in Eq. \eqref{decomposition} and defined in the text for the special case $L=6$ sites, $M=4$ fermions and $N=4$ colors.
          }
        \label{table_dec}
\end{table*}

Note that the dimension of the full Hilbert space $\mathcal{H}^N_L=\underset{M}{\oplus} \mathcal{H}^{M,N}_L$ is:
\begin{equation}
\sum_{M=0}^{M=NL}D^{M,N}_L= \sum_{q_L=0}^N \sum_{q_{L-1}=0}^N \cdots \sum_{q_1=0}^N  \prod_{j=1}^L \binom{N}{q_j}=2^{NL},
\end{equation}
by simple (iterative) application of the binomial formula.

The decomposition of the Hilbert space $\mathcal{H}^{M,N}_L$ into independent sectors invariant under the SU($N$) FHM (cf also Eq. 1 of the main text):
\begin{equation}
\label{supp_Hamiltonian}
H= \sum_{\langle i,j \rangle}  \Big{(}  -t_{ij} E_{ij}+ \text{h.c} \Big{)} + \frac{U}{2}  \sum_{i=1}^L E_{ii}^2, 
\end{equation}
where the $\mathrm{SU}(N)$ fermionic hopping terms $E_{ij}$ are 
\begin{equation}
\label{hopping}
E_{ij}=E_{ji}^{\dag}=\sum_{\sigma=1}^N c^{\dag}_{i \sigma}c_{j \sigma},
\end{equation}
 can be made following Eq. \eqref{decomposition}.
 The indices $i$ and $j$ in $E_{ij}$ are the {\bf site indices}.
 Note that $E_{ii}$ is the number of fermions on site $i$ (for $i=1 \cdots L$).
 
In fact since the $\mathrm{SU}(N)$ fermionic hopping terms $E_{ij}$ satisfy the U($L$) commutation relations (cf Eq. \eqref{supp_commutation}), $\mathcal{H}^{M,N}_L$ is also a representation of U($L$), but a {\it reducible} one,
and the dimensions  $d^{\alpha}_N$  appear as multiplicities according to $\mathcal{H}^{M,N}_L \cong \underset{\alpha}{\oplus} d^{\alpha}_N \mathcal{H}^{\bar{\alpha}}_L$,
where $\mathcal{H}^{\bar{\alpha}}_L$ is the irrep $\bar{\alpha}$ of U($L$). Each sector isomorphic to $\mathcal{H}^{\bar{\alpha}}_L$ gathers many-body wave-functions that belong
to the same $\mathrm{SU}(N)$ irrep $\alpha$ and is invariant under the action of the Hamiltonian H. 

To see this, one first needs to review the concept of {\it Highest Weight State} ($\vert \text{hws} \rangle $). For a given U($L$) irrep $\bar{\alpha}$, it is represented by the shape $\bar{\alpha}$ filled with $1$ for the first row (of length $\bar{\alpha}_1$), with $2$ for the second row (of length $\bar{\alpha}_2$), etc...(cf Fig. \ref{young_diagram}).
It is defined by the following properties: $E_{ii} \vert \text{hws} \rangle=\bar{\alpha}_i \vert \text{hws} \rangle$  $\forall i \in \llbracket 1;L \rrbracket$  and  $E_{ij} \vert \text{hws} \rangle=0$ for $i<j$ \cite{Paldus_2021}.
Crucially, in  $\mathcal{H}^{M,N}_L$, there are $d^{\alpha}_N$ orthonormal states $\vert \phi^{\text{hws}}_{\alpha, q}\rangle$ ($q=1 \cdots d^{\alpha}_N$) which have these properties and can then be represented by the same ssYT associated with the $\vert \text{hws} \rangle$ (cf Fig. 1 of the main text and the next subsection).

For example, when the number of particles $M$ is a multiple of $N$, the $\mathrm{SU}(N)$ singlets irrep is labelled by the perfectly rectangular $N-$rows Young diagram $\alpha=\alpha_{\mathcal{S},M}\equiv [M/N,M/N,...,M/N]$. In this case, that we will focus on in the next subsection, $d^{\alpha_{\mathcal{S},M}}_N=1$: there is only one state $\vert \phi^{\text{hws}}_{\alpha, 1}\rangle\equiv \vert \phi^{\text{hws}}_{\alpha_{\mathcal{S},M}}\rangle$ and it is the product of $\mathrm{SU}(N)$ singlets for sites $1, 2, \cdots M/N$, with no particles on sites $M/N+1, \cdots L$ (cf Fig. 1 of the main text and Fig. \ref{singlets_su2} below).

Secondly,  introducing the $\mathrm{SU}(N)$ {\bf total}  generators :
\begin{equation}
\label{total_Qr}
\mathcal{Q}_{r}=\sum_{i=1}^L \sum_{\sigma, \sigma'=1}^N c_{i,\sigma'}^{\dag} \lambda^{r}_{\sigma' \sigma}  c_{i,\sigma} \hspace{2cm} \text{for}  \hspace{1cm}   r=1 \cdots N^2-1
\end{equation}
(with the Gell-Mann matrices $(\lambda^{r}_{\sigma' \sigma} )$ introduced in the previous section), it appears that the states $\vert \phi^{\text{hws}}_{\alpha, q}\rangle$ are eigenvectors of the $N-1$ diagonal  $\mathcal{Q}_{r}$ (spanning the Cartan subalgebra) with different set of $N-1$ eigenvalues.
Then, since
 \begin{equation} 
 \label{commutation_Q_E}
 [\mathcal{Q}_{r}, E_{ij}]=0,  \,\,\, \forall  1 \leq i,j \leq L\,\,\text{and}\,\, r \in \llbracket 1; N^2-1\rrbracket,
 \end{equation}
the $d^{\alpha}_N$ sectors generated by the (repeated) applications of the $E_{ij}$ (for $1 \leq i,j \leq L$) on each $\vert \phi^{\text{hws}}_{\alpha, q}\rangle$ (for $q \in \llbracket 1; d^{\alpha}_N\rrbracket $) will be invariant under the action of $H$ in Eq. \eqref{supp_Hamiltonian}. Moreover, each sector will independently {\it represent} the $\mathrm{U}(L)$ irrep $\bar{\alpha}$, i.e being in one to one correspondance with the set of ssYT of shape $\bar{\alpha}$.
Finally, the total Casimirs of $\mathrm{SU}(N)$ of rank $1$ to $N$, which are polynomials in the $\mathcal{Q}_{r}$ \cite{Paldus_2021}, will also commute with the $E_{ij}$ from Eq \eqref{commutation_Q_E}, 
so that all the many-body states living in one such a sector will have the same features under any $\mathrm{SU}(N)$ {\bf total} transformation, i.e they will correspond to the same $\mathrm{SU}(N)$ irrep $\alpha$.

\subsection{Quadratic Casimir of SU($N$) and SU($N$) singlets}
\label{singlets}
Among the polynomial invariant operators, i.e the Casimirs, of U($N$),  the two simplest ones are the linear and the quadratic one:
\begin{align}
I_1&=\sum_i E_{ii}, \\
I_2&=\sum_{i,j} E_{ij}E_{ji},
\end{align}
which commute with all the generators $E_{ij}$  (for $1\leq i,j \leq N$) of U($N$), as a simple consequence of the commutation rules in Eq. (\ref{commutation}) \cite{Paldus_2021}.
From Schur Lemma, on a given irrep $\beta=[\beta_1,\beta_2,\cdots,\beta_N]$ of U($N$), they take constant values, which are given by \cite{Paldus_2021}:
\begin{align}
\chi (I_1)&=\sum_i \beta_{i}=M, \\
\chi (I_2)&=\sum_i \beta_{i}^2-\sum_j \bar{\beta_{j}}^2+NM,
\end{align}
where $M$ is the total number of boxes of the Young Tableau $\beta$ (and the number of particles in our system).
The quadratic Casimir of SU($N$) that we call $C_2$, is a quadratic polynomial in the SU($N$) generators, and a linear combination of $I_1$ and $I_2$ so that the commutation with the SU($N$) generators
can also be seen as a simple consequence of the properties of the U($N$) invariants.
One simple choice for $C_2$ is:
\begin{equation}
\label{C_2_def}
C_2=\frac{1}{2} \sum_{r} \mathcal{Q}_r^2=\sum_{i=1}^{N} H_{i}^2+\sum_{i\neq j} E_{ij}E_{ji}=I_2-I_1^2/N,
\end{equation}
so that the constant values on an irrep $\beta=[\beta_1,\beta_2,\cdots,\beta_N]$ of SU($N$) is:
\begin{equation}
\label{formula_C_2}
\chi (C_2)=\sum_i \beta_{i}^2-\sum_j \bar{\beta_{j}}^2+NM-M^2/N,
\end{equation}
in agreement with \cite{pilch_formulas_1984}.
Note that for the x-axis of Fig. 3 of the main text, we used a rescaled value of the quadratic Casimir: $\chi (C_2)/(2N)$, in order
to have the lowest non vanishing quadratic Casimir equal to $1$.

\begin{figure}[t]
    \centering
    \includegraphics[width=1\textwidth]{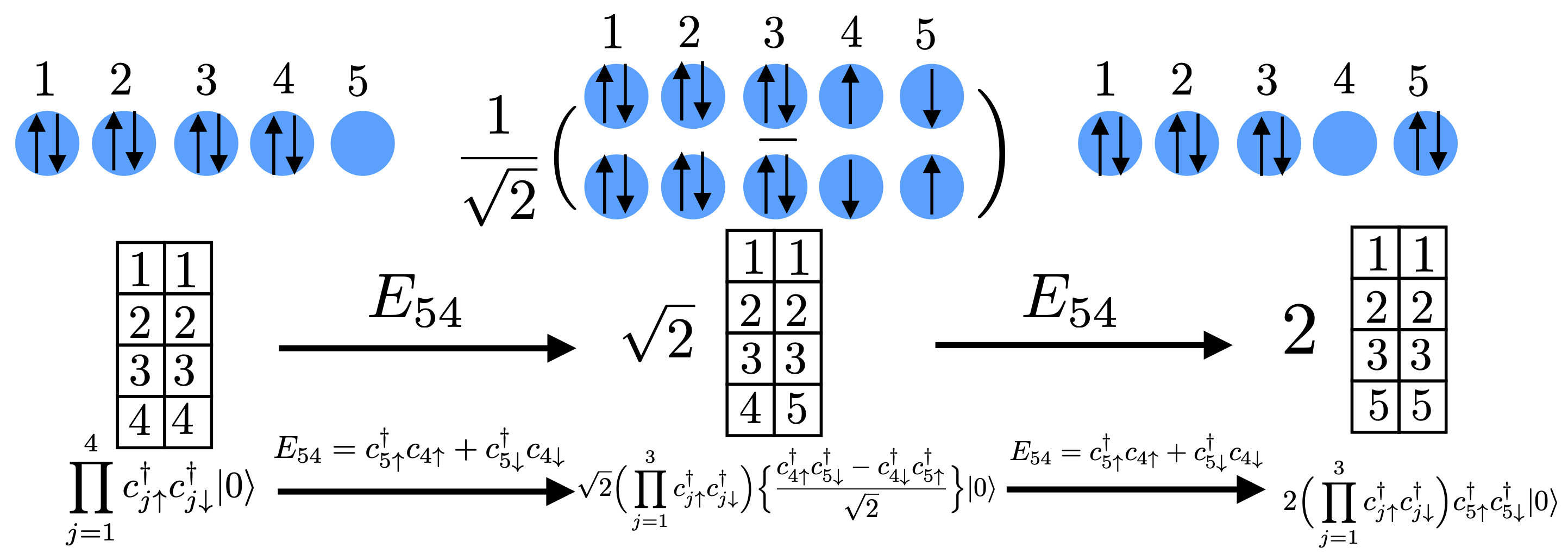}
    \caption{From top to bottom: three different ways to represent three different 8-fermions SU($2$) singlets states on a 5-sites lattice, i.e states living in the SU($2$) singlets subspace of $\mathcal{H}^{8,2}_5$ (cf text for details). Top row: sketch with up/down spins located over the 5 sites; Middle row: semi-standard Young Tableaux: $M=8$ boxes, $N=2$ columns, $M/N=4$ rows. The shape $\alpha_{\mathcal{S},M}=[2,2,2,2]$ is a rectangle with $N$ columns.
    Bottom row: corresponding states written in term of fermionic creation operators.
    At the very left, one has the Highest Weight State  $\vert \text{hws} \rangle \equiv \vert\phi^{\text{hws}}_{\alpha_{\mathcal{S},M}}\rangle$ for this sector.
    We pass from the left (resp. middle) state to the middle (resp. right) one by applying the SU($2$) invariant hopping operator $E_{54}=c^{\dag}_{5 \uparrow}c_{4 \uparrow}+c^{\dag}_{5 \downarrow}c_{4 \downarrow}$: the coefficients amplitudes $\sqrt{2}$ and $2=\sqrt{2}\times\sqrt{2}$, only present in the middle and bottom rows,
    can be calculated directly from the Gelfand Tsetlin rules (cf Eq. \eqref{Epmoins1p_rule}, Eq. \eqref{Eppmoins1_rule}, \eqref{mat_coeff} and \eqref{mat_coeff2}).}
    \label{singlets_su2}
\end{figure}

By definition, the SU($N$) singlets many-body states are invariant under the SU($N$) total transformation which are exponential of linear combinations of the total SU($N$) generators $\mathcal{Q}_{r}$ shown in Eq. \eqref{total_Qr}. Thus they should be cancelled out by any $\mathcal{Q}_{r}$ ($r=1 \cdots N^2-1$), and thus by the operator $C_2$ (when we use the total $\mathcal{Q}_{r}$ into Eq. \eqref{C_2_def}).
In fact, when $M$ is a multiple of $N$, i.e $M=pN$, with $p$ an integer, we can construct the very simple state introduced above, i.e 
\begin{equation}
\vert \phi^{\text{hws}}_{\alpha_{\mathcal{S},M}}\rangle= \Pi_{j=1}^{p} \Pi_{\sigma=1}^N c^{\dag}_{j \sigma} \vert 0 \rangle 
\end{equation}
 which is the only state (up to a phase) with one fermion from every specie on each site from  $1$ to $p=M/N$ with no fermion elsewhere (cf Fig. 1 of the main text and Fig. \ref{singlets_su2}).
 Physically speaking, you basically fill ({\it up to saturation}) the $p$ first sites of the lattice with SU($N$) colored fermions.
 Thus, it is the only state which satisfies:
 \begin{align}
 E_{ii}&\vert \phi^{\text{hws}}_{\alpha_{\mathcal{S},M}}\rangle=N \vert \phi^{\text{hws}}_{\alpha_{\mathcal{S},M}}\rangle &\text{for} \hspace{0.5cm} i \leq p, \\
 E_{ii}&\vert \phi^{\text{hws}}_{\alpha_{\mathcal{S},M}}\rangle= 0  &\text{for} \hspace{0.5cm} i > p, \\
 E_{ij} &\vert \phi^{\text{hws}}_{\alpha_{\mathcal{S},M}}\rangle =0 &\text{for} \hspace{0.5cm} i<j,
 \end{align}
   with $E_{ij}$ the SU($N$) invariant hopping terms defined in Eq. \eqref{hopping}.
   It also satisfies $\mathcal{Q}_{r}\vert \phi^{\text{hws}}_{\alpha_{\mathcal{S},M}}\rangle= 0$ (for $r=1 \cdots N^2-1$), so that is is a SU($N$) singlet.
   Moreover, using the commutation relation in Eq. \eqref{commutation_Q_E}, all the sector  generated by the (repeated) applications of the $E_{ij}$ will be the SU($N$) singlet subspace.
   One can also chek that $\chi (C_2)$ (cf Eq. \eqref{formula_C_2}) vanishes for the perfectly rectangular N-rows and p-columns shape $\alpha_{\mathcal{S},M}=[p,p,\cdots,p]$.
   
   Finally, in Fig. \ref{singlets_su2} we provide the correspondence between some SU($N$) singlets written with fermionic creation operators and with ssYT, 
   as well as the effect of some hopping operators $E_{ij}$ to illustrate the use of the Gelfand-Tsetlin matrix coefficients.
   
\subsection{The large U limit and the connection with the protocol for the Heisenberg model}
\label{largeU}
Let's mention the large $U$ limit at filling $1/N$.
 Then, the ssYT with strictly more than one occurence of a site index are too high in energy,
 and we are left with the subset of Standard Young Tableaux (no repetition allowed), in agreement with the ED protocol developed for the  HM~\cite{nataf_exact_2014}.
 Then, for each SYT $\vert \mu \rangle$ and for fixed $i=2\cdots L$, $a^j_{i-1}$ is either 0, either $a^j_{i-1}=\sqrt{\vert d-1\vert/d}$, after many cancellations between the numerator and the denominator in Eq. \eqref{mat_coeff}, where $j$ is the unique row where $p$ stands in $\vert \mu \rangle$,
 and with $d$ the Manhattan distance between $p-1$ and $p$. 
 Such a distance is defined by the number of horizontal steps (counted positively from right to left) in the row of $p-1$ plus the number of vertical steps (counted positively from top to bottom) in the column of
 $p$ to go from $p-1$ to $p$ in a given SYT $\vert \mu \rangle$.
 For instance, in the following SYT:
 \begin{equation}
 \ytableaushort{134,267,589}
 \end{equation}
 the Manhattan distance between $4$ and $5$ is equal to $+4$, while the Manhattan distance between $8$ and $9$ is $-1$.
 Then, applying the rules explained above in Eq. \eqref{Epmoins1p_rule} and in Eq. (4) of the main text, for a SYT such as:
  \begin{equation}
  \label{syt_typique}
 \ytableaushort{ \cdots \cdots q \cdots ,\cdots  \cdots \cdots  \cdots, \cdots {\scriptstyle q+1} \cdots, \cdots \cdots},
 \end{equation}
 one has:
 \begin{align}
 \label{eq_ssYT_1}
 \ytableausetup{boxsize=1.5em}
 E_{q q+1}\,\raisebox{1.8ex}{$\ytableaushort{ {\cdots} {\cdots} q {\cdots} ,{\cdots}  {\cdots} {\cdots}  {\cdots}, {\cdots} {\scriptstyle q+1}{\cdots}, {\cdots} {\cdots}}$}=\sqrt{\frac{\vert d-1\vert}{\vert d \vert }} \,\raisebox{1.8ex}{$\ytableaushort{ {\cdots} {\cdots} q {\cdots} ,{\cdots}  {\cdots} {\cdots}  {\cdots}, {\cdots} q {\cdots}, {\cdots} {\cdots}}$}.
\end{align}

 Thus, one has for instance:

  \begin{align}
\label{eq_ssYT_2}
 \ytableausetup{boxsize=1.5em}
E_{4 5}\,\raisebox{1.8ex}{$\ytableaushort{134,267,589}$}=\sqrt{\frac{3}{4}} \,\raisebox{1.8ex}{$\ytableaushort{134,267,489}$} \hspace{1cm} \text{and} \hspace{1cm} 
E_{8 9}\,\raisebox{1.8ex}{$\ytableaushort{134,267,589}$}=\sqrt{2} \,\raisebox{1.8ex}{$\ytableaushort{134,267,488}$}.
\end{align}
 
 Moreover, for a ssYT such as:
  \begin{equation}
 \ytableaushort{ \cdots \cdots q \cdots ,\cdots  \cdots \cdots  \cdots, \cdots q \cdots, \cdots \cdots},
 \end{equation}
 where the boxes with dots do not contain $q$ nor $q+1$, then from the rules in Eq. \eqref{Eppmoins1_rule} and Eq \ref{mat_coeff2}), one has:
 \begin{align}
 \label{eq_ssYT_3}
 \ytableausetup{boxsize=1.5em}
 E_{q+1 q}\,\raisebox{1.8ex}{$\ytableaushort{ {\cdots} {\cdots} q {\cdots} ,{\cdots}  {\cdots} {\cdots}  {\cdots}, {\cdots} q {\cdots}, {\cdots} {\cdots}}$}=\sqrt{\frac{\vert d \vert+1}{\vert d \vert }} \,\raisebox{1.8ex}{$\ytableaushort{ {\cdots} {\cdots} {\scriptstyle q+1} {\cdots} ,{\cdots}  {\cdots} {\cdots}  {\cdots}, {\cdots} q {\cdots}, {\cdots} {\cdots}}$}+\sqrt{\frac{\vert d \vert-1}{\vert d \vert }} \,\raisebox{1.8ex}{$\ytableaushort{ {\cdots} {\cdots} q {\cdots} ,{\cdots}  {\cdots} {\cdots}  {\cdots}, {\cdots} {\scriptstyle q+1} {\cdots}, {\cdots} {\cdots}}$}.
\end{align}
 
 Thus for a SYT like in Eq. \eqref{syt_typique}, one has:
 \begin{align}
 \label{eq_ssYT_4}
 \ytableausetup{boxsize=1.5em}
 E_{q+1 q}  E_{q q+1} \,\raisebox{1.8ex}{$\ytableaushort{ {\cdots} {\cdots} q {\cdots} ,{\cdots}  {\cdots} {\cdots}  {\cdots}, {\cdots} {\scriptstyle q+1} {\cdots}, {\cdots} {\cdots}}$}=\sqrt{1-\frac{1}{d^2}} \,\raisebox{1.8ex}{$\ytableaushort{ {\cdots} {\cdots} {\scriptstyle q+1} {\cdots} ,{\cdots}  {\cdots} {\cdots}  {\cdots}, {\cdots} q {\cdots}, {\cdots} {\cdots}}$}+\Big{(}1-\frac{1}{d}\Big{)} \,\raisebox{1.8ex}{$\ytableaushort{ {\cdots} {\cdots} q {\cdots} ,{\cdots}  {\cdots} {\cdots}  {\cdots}, {\cdots} {\scriptstyle q+1} {\cdots}, {\cdots} {\cdots}}$},
\end{align}
where the first term of the RHS disappears when $d=\pm1$ (meaning $q$ and $q+1$ side by side in the same row/column).
 
Finally, the transposition $P_{q q+1}$ being equal to $-E_{q+1 q}E_{q q+1}+1$, we recover the Young rules for the orthogonal representations of the algebra of the group of permutations on the basis of SYTs in a given irrep $\alpha$~\cite{nataf_exact_2014,young_onsubstitutional_1900}.

\subsection{Sp($N$) singlets}
\label{symplectic}
Since $Sp(N)\subset SU(N) \subset U(N)$, the  $N(N+1)/2$  generators of Sp($N$) are linear superpositions of the SU($N$) generators and of the U($N$) generators.
So for $N=4$, the $10$ generators of Sp($N$) can be chosen as \cite{cahn_2014}:

\begin{align}
\label{spn_generators}
\tilde{E}_{+1}&=E_{12}-E_{43}, &\tilde{E}_{-1}&=E_{21}-E_{34},  \nonumber\\ \nonumber
\tilde{E}_{+2}&=E_{24},&\tilde{E}_{-2}&=E_{42}, \nonumber \\
\tilde{E}_{+3}&=E_{14}+E_{23}, &\tilde{E}_{-3}&=E_{41}+E_{32},\\
\tilde{E}_{+4}&=E_{13}, &\tilde{E}_{-4}&=E_{31}, \nonumber \\
\tilde{H}_{1}&=E_{11}-E_{33}-E_{22}+E_{44}, &\tilde{H}_{2}&=E_{22}-E_{44},  \nonumber
\end{align}

Then, each $SU(N)$ singlet is also invariant under Sp($N$) so that the dimension $d^{\text{Sing}}_{Sp(N), L}$ of the $Sp(N)$ singlets subspace should always be larger than
the dimension of the $SU(N)$ singlets subspace, $d^{\text{Sing}}_{SU(N), L}\equiv d^{\bar{\alpha}_{\mathcal{S},M}}_L$, no matter the filling $M$.
On the other hand, $d^{\text{Sing}}_{Sp(N), L}$ is also smaller than $d^{U(1)}_{L,M,N}$, i.e the dimension of the sector addressed in standard ED,
which gathers many-body states with a fixed and equal number of fermions of each color (conserving the $U(1)$ symmetry).
To quantitatively see {\it how smaller} it is, and thus how helpful would be the implementation of the Sp($N$) symmetry, we have calculated these dimensions for $N=4$ colors, $M=4$ fermions and for $L=2$ to $L=8$ sites.
On a {\it standard} fermionic Fock state basis of the sector $\mathcal{H}^{M,N}_L$ (i.e without the use of ssYT), we have implemented both the SU($N$) quadratic Casimir $C_2^{SU(N)}\equiv C_2$ (cf Eq. \eqref{C_2_def} above
with total SU($N$) generators, i.e from Eq. \eqref{total_Qr}), and the Sp($N$) quadratic Casimir $C_2^{Sp(N)}$, which reads for $N=4$ (follwing the general method exposed in \cite{cahn_2014}):
\begin{equation}
\label{c2sp4}
C_2^{Sp(4)}=\tilde{H}_1^2+2\tilde{H}_2^2+2\tilde{H}_1\tilde{H}_2+\tilde{E}_{+1}\tilde{E}_{-1}+\tilde{E}_{-1}\tilde{E}_{+1}+2(\tilde{E}_{+2}\tilde{E}_{-2}+\tilde{E}_{-2}\tilde{E}_{+2})+\tilde{E}_{+3}\tilde{E}_{-3}+\tilde{E}_{-3}\tilde{E}_{+3}+2(\tilde{E}_{+4}\tilde{E}_{-4}+\tilde{E}_{-4}\tilde{E}_{+4}),
\end{equation}
where we have used the total Sp($N$) generators (cf Eq. \eqref{spn_generators} with $E_{\mu \nu}=\sum_{i=1}^L c_{i,\mu}^{\dag}  c_{i,\nu}$ for $\mu,\nu=1 \cdots 4$) .

We have determined both $d^{\text{Sing}}_{SU(N), L}$ and $d^{\text{Sing}}_{Sp(N), L}$ by calculating the dimension of the null spaces of $C_2^{SU(4)}$ and of $C_2^{Sp(4)}$ to fill up the array \ref{table_dim},
which shows that $d^{\text{Sing}}_{Sp(N), L}$ is always larger than $d^{\text{Sing}}_{SU(N), L}$ for $N=4$ colors and $M=4$ fermions, but still much smaller than $d^{U(1)}_{L,M=4,N=4}=L^4$ which is the dimension of the sector
where each color is equally represented on the lattice.

\begin{table*}[ht]
        \centering
        \begin{tabular}{|c|c|c|c|c|c|c|c|}
          \hline
            
                L&2&3&4&5&6&7&8 \\
                 \hline
  $d^{U(1)}_{L,M=4,N=4}=L^4$&16&81&256&625&1296&2401&4096 \\

                \hline
               $d^{\text{Sing}}_{SU(N=4), L}$&5&15&35&70&126&210&330 \\
              \hline
   $d^{\text{Sing}}_{Sp(N=4), L}$&6&21&55&120&231&406&666 \\
\hline

        \end{tabular}
        \caption{Dimensions of some relevant sectors included in the Hilbert space $\mathcal{H}^{M,N}_L$ for $N=M=4$ as a function of $L$.
        Top row: $d^{U(1)}_{L,M,N}=\binom{L}{M/N}^N=L^4$ is the dimension of the sector usually considered in standard ED calculation, with a fixed and equal number of fermions from each color,
        i.e when the conserved U($1$) symmetry is implemented; Middle row: $d^{\text{Sing}}_{SU(N=4), L}$, dimension of the SU($4$) singlets subspace, obtained by calculating the dimension of the null space
        of  $C_2^{SU(4)}$ (or by using Eq.~\eqref{supp_equation_dim}); Bottom row: $d^{\text{Sing}}_{Sp(N=4), L}$, dimension of the Sp($4$) singlets subspace, obtained by calculating the dimension of the null space
        of  $C_2^{Sp(4)}$ (cf Eq. \eqref{c2sp4}). CF text for details.
          }
        \label{table_dim}
\end{table*}

\section{Complementary numerical results}
\label{sec: num}
\subsection{SU(5) on a square lattice}
\label{sec: su5}
\begin{figure}[t]
    \centering
    \includegraphics[width=1.0\textwidth]{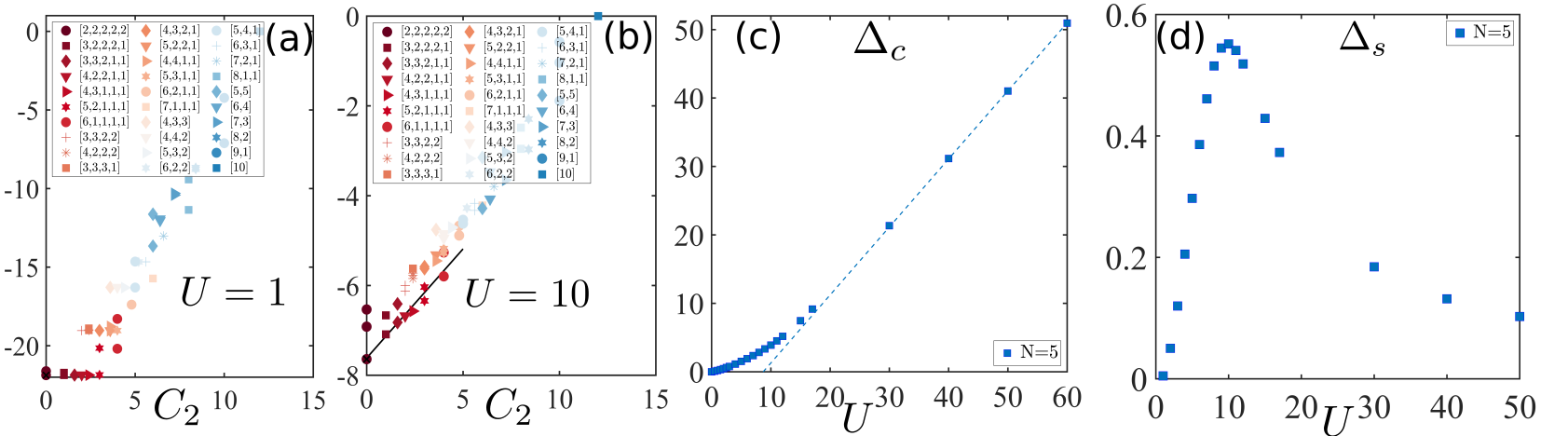}
    \caption{Energy spectra of the FHM in Eq.~(1) (main text) (with $t_{ij}=1$
at filling $1/N$) as a function of the quadratic casimir $C_2$ for $N=5$, for the L=10 periodic squared lattice for various values of $U$, (a) $U=1$, (b) $U=10$. In (b), the black line is a guide for the eye to show the presence of the Anderson tower of states.
(c) Charge gap for the FHM on the $L=10$  squared periodic lattice as a function of the one-site repulsion $U$. The dashed line (fit of the form $\Delta_{\rm{c}} = a*U -b$) crosses the $x$ axis at $U_c \approx 8.75$,  separating the small U metallic phase from the Mott insulators. (d) Spin gaps as a function of $U$. It exhibits a peak at $U \approx 8$.}
    \label{fig:su5}
\end{figure}
 In this section, we show the data for SU($5$) on a square lattice with L=10 sites at filling $1/5$, that we have discussed  in the main text. The long-range color ordered state with broken $SU(5)$ symmetry is revealed by an Anderson tower of states (Atos), clearly visible on Fig.~\ref{fig:su5} (b)  for $U=10$.  The situation is quite different at small $U$, as seen from Fig.~\ref{fig:su5} (a), where the Atos has disappeared and an energy plateau appears. In Fig.~\ref{fig:su5} (c), we plot the charge gap $\Delta_c$ to locate the boundary $U=U_c$ between the metallic and the LRO phases. We see that $\Delta_{\rm{c}}$
 opens rapidly around $U \approx 8$.
 
 To estimate the critical point roughly, we use a fit of the form $\Delta_{\rm{c}} = a*U -b$. The latter corresponds to the expected behavior of the charge gap at large U, since the energy cost to add one particle should be equal to $U$: using the definition of $\Delta_{\rm{c}}$ one can readily show that for $U\to \infty$, $\Delta_{\rm{c}} = U$. Then, such a linear function shown in dashed lines in the figures cross the x axis at $U=U_{\rm{c}}$.
 However, due to the small curvature of $\Delta_{\rm{c}}$, it depends on the values of $U$ used for the fit; we took different sets belonging to $U \in [40; 100]$.
 With this method, we obtain $8.6 < U_{\rm{c}} < 8.9 $  with $ 0.992 < a < 0.997$. 
 In addition, in Fig.~\ref{fig:su5} (d), we show the spin gap, defined as the difference between the SU($5$) adjoint and the SU($5$) singlet minimal energies. The spin gap increases rapidly in the metallic phase, reaching its maximal value around $U\approx8$, close to  $U=U_{\rm{c}}$,
 a feature also observed in the other cases studied here.

\subsection{SU(2) on the  triangular lattices with L=12 and L=16 sites}
\label{sec: su2}
\begin{figure}[t]
    \centering
    \includegraphics[width=1.0\textwidth]{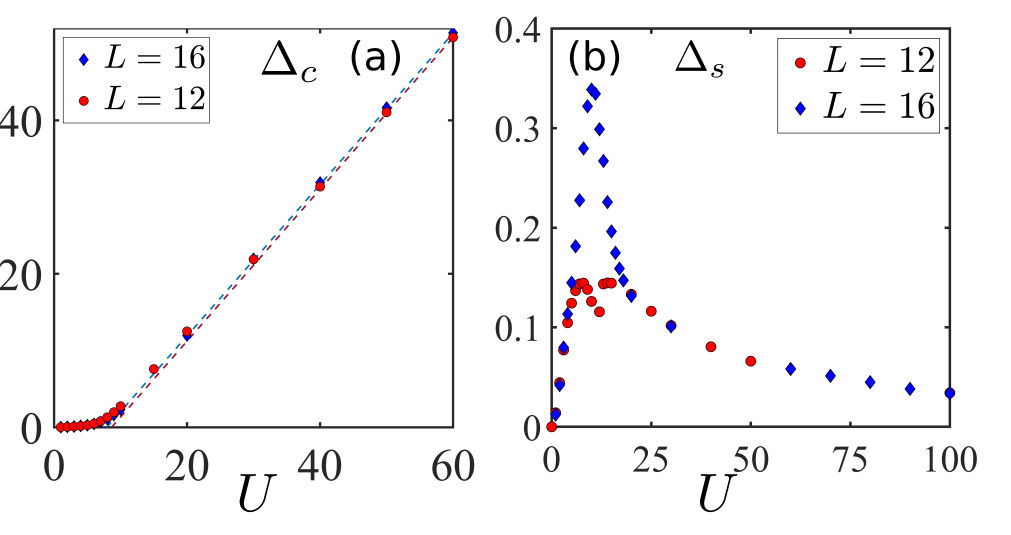}
    \caption{(a) Charge gaps for the FHM on the $L=12$ and $L=16$ sites triangular lattice for N=2. The dashed lines correspond to a fit of the form $\Delta_{\rm{c}}= a*U - b$. They cross the x axis at $U=U_{\rm{c}}=8.6 \pm 0.3$ for $L=12$ and at $U=U_{\rm{c}}=7.9 \pm 0.3$ $L=16$, separating the small $U$ metallic phase from the Mott insulators. (b) Spin gaps for FHM on the $L=12$ and $L=16$ sites triangular lattice for N=2. A peak is present at $U \simeq 8$ in both cases.}
    \label{fig:su2}
\end{figure}

In this section, we analyze the case SU($2$) on a triangular lattice with $L=12$ and $L=16$ sites at filling $1/2$. In Fig.~\ref{sec: su5} (a), we show the charge gap as a function of $U$ for $L=12$ (blue diamond) and $L=16$ (red circle). Using the fitting procedure detailed in the Sec.~\ref{fig:su5} above, we estimate the phase boundary between a metallic phase and an insulating phase: we have  $U_{\rm{c}}=8.6 \pm 0.3$ for $L=12$ and $U_{\rm{c}}=7.9 \pm 0.3$ for $L=16$. In Fig.~\ref{sec: su5} (b), we show the spin gap as a function of U. 

\subsection{Evolution of correlations with U}
\label{sec: correlation}

\begin{figure}[t]
    \centering
    \includegraphics[width=1.0\textwidth]{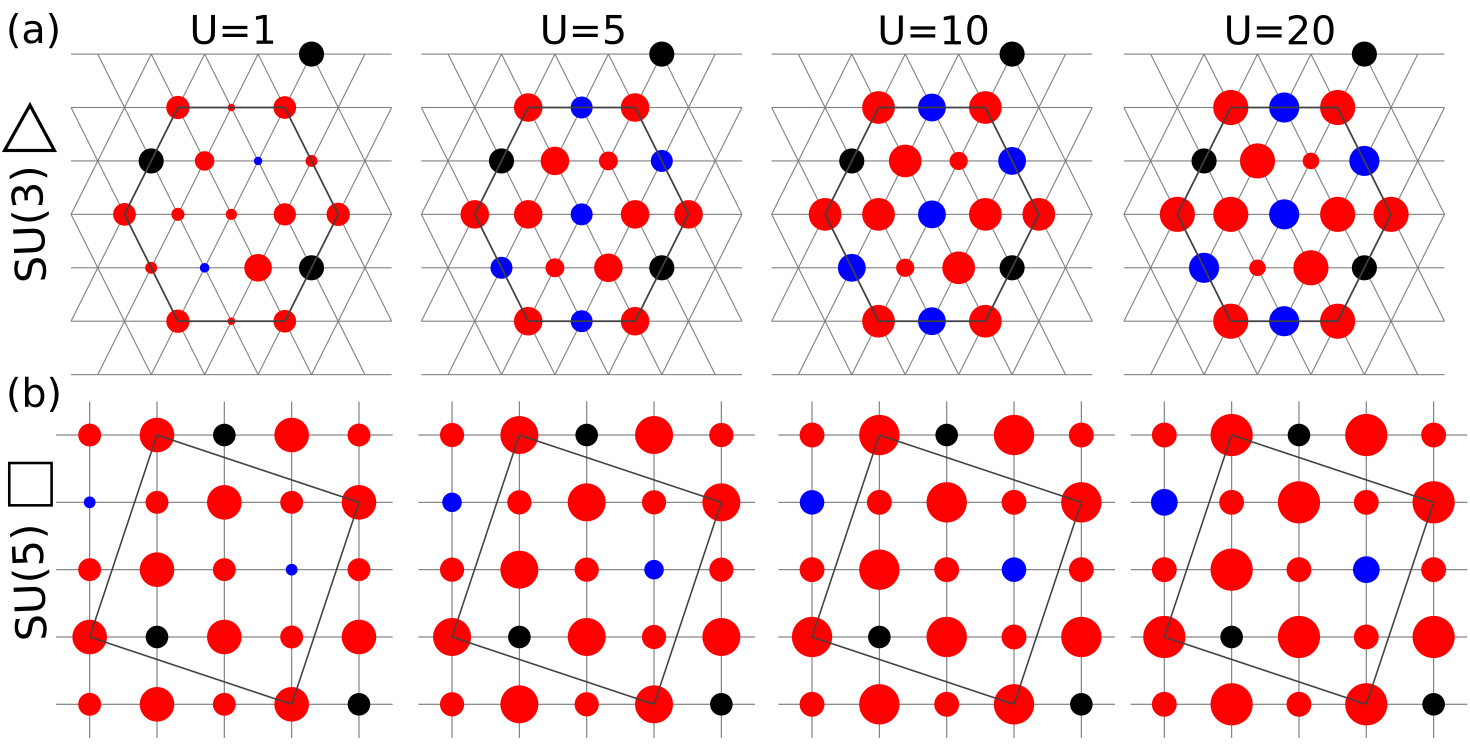}
    \caption{Evolution of the correlations patterns of the ground states of the FHM at filling 1/N for various from left to right $U=1, 5, 10, 20$. (a) On a triangular lattice SU($3$). (b) On a  square lattice SU($5$). They are defined as $\langle P_{1j}\rangle -1/N$ where $P_{1j}\equiv -1+E_{1j}E_{j1}$, with the reference site  $1$ in black, and $j$ the site indices being blue (resp. red) for positive (resp. negative) correlation, with area proportional to its absolute value.}
    \label{fig: correlation}
\end{figure}

In this section, we study how the real space correlations, defined as  $\langle P_{1j}\rangle -1/N$ where $P_{1j}\equiv -1+E_{1j}E_{j1}$, evolve while increasing the on-site interaction U. In Fig.~\ref{fig: correlation}, we show the correlations for two lattice geometries and different U. From left to right, we have U=1, U=5, U=10, and U=20. In (a), we consider the case SU($3$) on a triangular lattice. We see that for $U=1$, the correlations do not display the three-sublattice Neel order. The order becomes clearly visible at $U=5$. In (b), the case SU($5$) on a square lattice is shown. Here, the pattern compatible with LRO is visible at any $U$, although with smaller correlation strengths for small $U$. Note that while increasing the on-site interaction $U$, we do not observe the appearance of a new ( or differently configured) long-range orders. For completeness, we have checked that patterns do not evolve for $U$ up to $50$ (not shown).


\end{document}